\documentclass[journal=jpc,manuscript=article]{achemso}

\usepackage{chemformula} 
\usepackage[T1]{fontenc} 
\usepackage{physics}
\usepackage{multirow}
\usepackage{caption}
\usepackage{subcaption}
\usepackage{modiagram}
\usepackage{tikz}
\usetikzlibrary{decorations.pathmorphing}
\usepackage{bm}
\usepackage{graphicx,graphics}

\newcommand*{\citen}[1]{%
  \begingroup
    \romannumeral-`\x 
    \setcitestyle{numbers}%
    \cite{#1}%
  \endgroup
}
\tikzset{
    ncbar angle/.initial=90,
    ncbar/.style={
        to path=(\tikztostart)
        -- ($(\tikztostart)!#1!\pgfkeysvalueof{/tikz/ncbar angle}:(\tikztotarget)$)
        -- ($(\tikztotarget)!($(\tikztostart)!#1!\pgfkeysvalueof{/tikz/ncbar angle}:(\tikztotarget)$)!\pgfkeysvalueof{/tikz/ncbar angle}:(\tikztostart)$)
        -- (\tikztotarget)
    },
    ncbar/.default=0.5cm,
}

\tikzset{square left brace/.style={ncbar=0.1cm}}
\tikzset{square right brace/.style={ncbar=-0.1cm}}

\tikzset{round left paren/.style={ncbar=0.5cm,out=120,in=-120}}
\tikzset{round right paren/.style={ncbar=0.5cm,out=60,in=-60}}

\SectionNumbersOn
\author{Dilhan Manawadu}
\affiliation[]
{Department of Chemistry, Physical and Theoretical Chemistry Laboratory, \\University of Oxford, Oxford, OX1 3QZ, United Kingdom}
\alsoaffiliation[Second University]
{Linacre College, University of Oxford, Oxford, OX1 3JA, United Kingdom}
\email{dilhan.manawadu@chem.ox.ac.uk}
\author{Timothy N.\ Georges}
\affiliation[]
{Department of Chemistry, Physical and Theoretical Chemistry Laboratory, \\University of Oxford, Oxford, OX1 3QZ, United Kingdom}
\alsoaffiliation[Second University]
{Brasenose College, University of Oxford, Oxford, OX1 4AJ, United Kingdom}
\author{William Barford}
\affiliation[]
{Department of Chemistry, Physical and Theoretical Chemistry Laboratory, \\University of Oxford, Oxford, OX1 3QZ, United Kingdom}
\email{william.barford@chem.ox.ac.uk}

\graphicspath{{./figures/}}
\title[An \textsf{achemso} demo]
  {Photoexcited state dynamics and \\singlet fission in carotenoids}

\abbreviations{IR,NMR,UV}

\begin{document}

\begin{tocentry}
\resizebox{8.4cm}{4.5cm}{%
\begin{tikzpicture}
	\coordinate (origin) at (0,3);
	\coordinate (x) at (8,3);			
	\coordinate (y) at (0,10);
	
	\draw [->,black,very thick] (1.5,3) -- (1.5,7.25) ;
	\node[align=center,font=\large] at (1.25,5) {$\hat{\mu}$};
	
	\draw [->,red,very thick] (origin) -- (x) ;
	\draw [->,red,very thick] (origin) -- (y) ;
	
	 \node[align=center] at (4,2.75) {time};
	 \node[align=center,rotate=90] at (-0.25,6.5) {Energy};
	
	 \draw [black,fill=yellow,thick] (1.5,8) ellipse (1cm and 0.75cm);
	\node[align=center,font=\small] at (1.25,8) {$\bm{+}$};
	\node[align=center,font=\small] at (1.75,8) {$\bm{-}$};

	\draw [blue, very thick] (2.25,7.5) edge[bend right, ] (5,5) ;
	
	 \draw [black,fill=yellow,thick] (5.5,5) ellipse (0.65cm and 0.375cm);
	\node[align=center,font=\small] at (5.25,5) {$\bm{+}$};
	\node[align=center,font=\small] at (5.75,5) {$\bm{-}$};
	
	\node[align=center,font=\footnotesize] at (6.32,5) {$+$};
	
	\draw [black,fill=green,thick] (7.15,5) ellipse (0.65cm and 0.375cm);
	\draw [->,thick] (6.75,4.75) -- (6.75,5.25) ;
	\draw [->,thick] (6.95,4.75) -- (6.95,5.25) ;
	
	\draw [->,thick]  (7.35,5.25) -- (7.35,4.75) ;
	\draw [->,thick]  (7.55,5.25) -- (7.55,4.75);
	
	\draw[draw=black,thick] (4.75,4.25) rectangle ++ (3.25,1.5);
	
	\node[align=center,font=\small] at (1.5,9.25) {$1^1 B_u^+$};
	\node[align=center,font=\small] at (6.25,6.25) {$2^1 A_g^-$};
	
\end{tikzpicture}
}
%
%
%

\end{tocentry}

\begin{abstract}

We describe our dynamical simulations of the excited states of the carotenoid, neurosporene, following its photoexcitation into the `bright' (nominally $1^1B_u^+$) state. We employ the adaptive tDMRG method on the UV model of $\pi$-conjugated electrons and use the Ehrenfest equations of motion to simulate the coupled nuclei dynamics. To account for the experimental and theoretical uncertainty in the relative energetic ordering of the nominal $1^1B_u^+$ and $2^1A_g^-$  states at the Franck-Condon point, we consider two parameter sets. In both cases there is ultrafast internal conversion from the `bright' state to a `dark' singlet triplet-pair state, i.e., to one member of the `$2A_g$' family of states.
For one parameter set internal conversion from the $1^1B_u^+$ to $2^1A_g^-$ states occurs via the dark, intermediate $1^1B_u^-$ state. In this case there is a crossover of the $1^1B_u^+$ and $1^1B_u^-$ diabatic energies within 5 fs and an associated avoided crossing of the $S_2$ and $S_3$ adiabatic energies. After the adiabatic evolution of the $S_2$ state from predominately $1^1B_u^+$ character to predominately $1^1B_u^-$ character, there is a slower nonadiabatic transition from $S_2$ to $S_1$, accompanied by an increase in the population of the $2^1A_g^-$  state. For the other parameter set the $2^1A_g^-$ energy lies higher than the $1^1B_u^+$ energy at the Franck-Condon point. In this case there is crossover of the  $2^1A_g^-$ and  $1^1B_u^+$ energies and an avoided crossing of the $S_1$ and $S_2$ energies, as the $S_1$ state evolves adiabatically from being of $1^1B_u^+$ character to $2^1A_g^-$ character.
We make a direct connection from our predictions to experimental observables by calculating the transient absorption. For the case of direct $1^1B_u^+$ to $2^1A_g^-$ internal conversion, we show that the dominant transition at ca.\ 2 eV, being close to but lower in energy than the $T_1$ to $T_1^*$ transition, can be attributed to the  $2^1A_g^-$ component of $S_1$. Moreover, we show that it is the charge-transfer exciton component of the  $2^1A_g^-$ state that is responsible for this transition (to a higher-lying exciton state), and not its triplet-pair component.
We next discuss the microscopic mechanism of `bright' to `dark' state internal conversion, emphasising that this occurs via the exciton components of both states.
Finally, we describe a mechanism whereby the strongly bound intrachain triplet-pairs of the `dark' state may undergo interchain exothermic dissociation. This mechanism relies on the possibility of the unbound interchain triplets being energetically stabilized by quantum deconfinement, and larger bond and torsional reorganization energies. We predict that this is only possible if the molecules are twisted in their ground states.
The computational methodology underlying the calculations described here is explained in our companion paper, \textit{Dynamical simulations of carotenoid photoexcited states using density matrix renormalization group techniques}, D.\ Manawadu, D.\ J.\ Valentine, and W.\ Barford, \emph{J. Chem. Theo. Comp.} (2023).

\end{abstract}

\newpage
\section{Introduction}

Carotenoids are a class of linear polyenes of high natural abundance. Carotenoids found in photosynthetic systems serve the dual functions of enhancing their light harvesting properties by absorbing photons in the visible spectrum not accessible to chlorophylls and protecting the light harvesting complexes from excess light.\cite{Fraser2001,Uragami2020} The study of carotenoid photophysics is important  for understanding their functions in photosynthetic systems.\cite{Hashimoto2016}

The quasi-one-dimensional nature of polyenes enhances electron-electron interactions and electron-nuclear coupling, and gives rise to a complex excited state structure.\cite{Hudson1972,Schulten1972,Hayden86,Tavan1987,Chandross1999,Bursill1999,Barford01,Barford2013c}
In 1972, it was observed that in polyenes there exists a symmetry-forbidden $2^1 A_g^-$ `dark' excited state (usually labelled $S_1$) below the photoexcited $1^1 B_u^+$ state (usually labelled $S_2$).\cite{Hudson1972,Schulten1972} Then, in 1987, it was shown that there exists other dark states within the $S_2-S_1$ gap.\cite{Tavan1987} Upon photoexcitation to the $S_2$ state, these dark excited states are involved in the ultrafast internal conversion processes, giving rise to the exotic photophysical properties of polyenes, including singlet fission.

Singlet fission is a photophysical process by which a singlet photoexcited state dissociates into two spin uncorrelated triplets. In carotenoids, the first step of singlet fission is understood to be the internal conversion from the photoexcited $1^1 B_u^+$ state to a correlated singlet triplet-pair state. The mechanism of this internal conversion process has been heavily debated.\cite{Polivka2009}
Spectroscopic studies of carotenoid excited states reveal that although the $S_2 - S_1$ gap increases with conjugation length, the $S_2$ lifetime behaves non-monotonically: initially increasing and then decreasing with conjugation length, in an apparent violation of the energy gap law. This implies for longer carotenoids, as for polyenes, that an intermediate dark state exists   which might be involved in the internal conversion process.\cite{Frank1997,Kosumi2006}

Recent theoretical work using diabatic models continues to provide evidence for the importance of the low-lying dark excited states of polyenes to their photophysics, especially in the singlet fission process.\cite{Santra2022} However, \emph{ab-initio} calculations of polyene excited states argue that nuclear reorganization following photoexcitation can facilitate the internal conversion process, without needing to invoke intermediate dark states.\cite{Taffet2019e,Khokhlov2020b}

In a theoretical and computational study using the density matrix renormalization group (DMRG) method to solve the Pariser-Parr-Pople-Peierls (PPPP) model of $\pi$-conjugated electrons, Valentine \emph{et al}.\cite{Valentine20} showed that the dark excited states, $2^1 A_g^- ,1^1 B_u^- ,3^1 A_g^-,\dots$ belong to the same family of fundamental excitation with different center-of-mass kinetic energies. The triplet-pair nature of this $2A_g$ family (or band) of states was established by calculating the spin-spin correlation, bond dimerization, and triplet-pair overlaps.

In a recent paper we described our dynamical simulations of singlet triplet-pair production in photoexcited zeaxanthin using the adaptive time-dependent DMRG (tDMRG) method and Ehrenfest dynamics.\cite{Manawadu2022} We chose a parameter regime where at the Franck-Condon point the diabatic energies satisfy $E(2^1 A_g^-) < E(1^1 B_u^+) < E(1^1 B_u^-)$, while the adiabatic singlet states are, $S_1 \approx 2^1 A_g^-$,  $S_2 \approx 1^1 B_u^+$, and $S_3 \approx 1^1 B_u^-$. Within 5 fs of  photoexcitation to $S_2$ there is a diabatic crossover of the $1^1 B_u^+$ and $1^1 B_u^-$ energies, but an avoided crossing of the $S_2$ and $S_3$ energies, such that $S_2$ evolves quasiadiabatically from the $1^1 B_u^+$ state to the $1^1 B_u^-$ state. Since zeaxanthin possesses $C_{2}$ symmetry, there is no further interstate conversion from the $1^1 B_u^-$ to the $2^1 A_g^-$.

In this paper we extend that work to investigate internal conversion in neurosporene, a molecule of 18 conjugated carbon atoms that does not possess $C_{2}$ symmetry, thus permitting $1^1 B_u^+$ to $2^1 A_g^-$ state conversion. We consider two parameter sets, (a) $E(2^1 A_g^-) < E(1^1 B_u^+) < E(1^1 B_u^-)$ at the Franck-Condon point, where internal conversion from the $1^1 B_u^+$ to $2^1 A_g^-$ states occurs via the intermediate $1^1 B_u^-$ state and (b) $E(1^1 B_u^+) < E(2^1 A_g^-) < E(1^1 B_u^-)$ at the Franck-Condon point, where there is direct internal conversion from the $1^1 B_u^+$ to $2^1 A_g^-$ states. In both cases we show that after 50 fs the yield of the singlet triplet-pair states is ca.\ 65 \%.

As well as describing the dynamical simulations of internal conversion, this paper has three further goals. First, we discuss our calculated transient absorption spectra from the evolving state and we use our results to interpret experimental observations. In particular, we argue that a dominant transient absorption feature at ca.\ 2 eV, close to but lower in energy than a triplet state absorption, originates from the charge-transfer exciton component  of the `dark', $2^1 A_g^-$ state. Second, using the theory of the `dark' state of ref \citenum{Barford2022}, where it was shown that this state contains both singlet triplet-pair and charge-transfer exciton character, we describe the microscopic mechanism of `bright' to `dark' state internal conversion in carotenoids. Finally, we examine the question of whether the bound intrachain triplet-pairs can undergo exothermic interchain dissociation. We show that this is possible if interchain transfer is accompanied by torsional relaxation, implying that the carotenoid dimers should have a twisted ground state geometry.

A companion paper\cite{Manawadu2022a} describes our computational DMRG methodology for simulating the excited state dynamics of strongly correlated electron systems. It also analyses the physics of the diabatic crossover and the adiabatic avoided crossing in greater detail.

The contents of this paper is the following. After briefly introducing our model in section 2, section 3 describes our dynamical simulations of photoexcited state interconversion, our calculation and interpretation of the transient absorption, and concludes with a microscopic explanation of the ultrafast `bright' to `dark' interstate conversion observed in carotenoids. Section 4 shows that interchain singlet dissociation into triplets can be exothermic if accompanied by torsional relaxation. We conclude in section 5.


\section{Computational methods}\label{sec:comp}

The $\pi$-electron system is described by the extended Hubbard (or UV) model, defined by
\begin{equation}
	\hat{H}_{\textrm{UV}} = -2\beta \sum_{n=1}^{N-1}  \hat{T}_n + U \sum_{n=1}^N  \big( \hat{N}_{n \uparrow} - \frac{1}{2} \big) \big( \hat{N}_{n \downarrow} - \frac{1}{2} \big)
+ \frac{1}{2} \sum_{n=1}^{N-1} V \big( \hat{N}_n -1 \big) \big( \hat{N}_{n+1} -1 \big), \label{eqn:UV}
\end{equation}
where $n$ labels the $n^\textrm{th}$\ C-atom, $N$ is the number of conjugated C-atoms and $N/2$ is the number of double bonds.
$\hat{T}_ n = \frac{1}{2}\sum_{\sigma} \left( \hat{c}^{\dag}_{n , \sigma} \hat{c}_{n + 1 , \sigma} + \hat{c}^{\dag}_{n+ 1 , \sigma} \hat{c}_{n  , \sigma} \right)$  is the bond order operator, $\hat{c}^{\dag}_{n , \sigma} (\hat{c}^{}_{n , \sigma})$ creates (destroys) an electron with spin $\sigma$ in the $p_z$ orbital of the $n^{\textrm{th}}$ C-atom, and $\hat{N}_n$ is the number operator. $U$ and $V$ correspond to Coulomb parameters which describe interactions of two electrons in the same orbital and nearest neighbors, respectively, and $\beta = 2.4$ eV represents the electron hopping integral between neighboring C-atoms.

The electron-nuclear coupling is described by
\begin{equation}
	\hat{H}_{\textrm{e-n}} = 2 \alpha \sum_{n=1}^{N-1}  \left( u_{n+1} - u_n \right)  \hat{T}_n, \label{eqn:e-n}
\end{equation}
where $\alpha=4.593$ eV \AA$^{-1}$ is the electron-nuclear coupling parameter and $u_n$ is the displacement of the $n^\textrm{th}$\ C-atom from its undimerized geometry.
The nuclear degrees of freedom are described by the classical Hamiltonian
\begin{equation}
	\hat{H}_{\textrm{elastic}} = \frac{K}{2} \sum_{n=1}^{N-1} \left( u_{n+1} - u_n \right)^2 , \label{eqn:elastic}
\end{equation}
where $K=46$ eV {\AA}$^{-2}$ is the nuclear spring constant.

To project out the high spin eigenstates of the Hamiltonian, we complement the Hamiltonian with
\begin{equation}
	\hat{H}_\lambda = \lambda \hat{S}^2,
\end{equation}	
where $\hat{S}$ is the total spin operator and $\lambda > 0$.

The UV-Peierls (UVP) Hamiltonian, defined by
\begin{equation}
	\hat{H}_{\textrm{UVP}} = \hat{H}_{\textrm{UV}}+\hat{H}_{\textrm{e-n}}+\hat{H}_{\textrm{elastic}},  \label{eqn:UVP}
\end{equation}
is invariant under a particle-hole transformation. For idealized carotenoid structures
with  two-fold rotation symmetry, its eigenstates will have definite $C_{2}$ and particle-hole symmetries.\cite{Barford2013c} Therefore its eigenstates are labelled either $A_g^\pm$ or $B_u^\pm$. We define the eigenstates of $(\hat{H}_{\textrm{UVP}} +  \hat{H}_\lambda)$ as diabatic states. The ordering of these states is highly sensitive to the $U$ and $V$ parameters. It was shown in ref \citen{Manawadu2022} that for the UVP model with $U=7.25$ eV and $V=2.75$ eV, the $1^1 B_u^-$ relaxed energy is lower than the $1^1 B_u^+$ relaxed energy for chain lengths $N \geq 10$, while the $1^1 B_u^-$ vertical energy is higher than the $1^1 B_u^+$ vertical energy for $N \leq 22$. Both the vertical and relaxed $2^1 A_g^-$ energies are below the $1^1 B_u^+$ energies for all relevant chain lengths. This implies that for these model parameters  internal conversion from the $1^1 B_u^+$ state to the $2^1 A_g^-$ state is expected to proceed \emph{via} the $1^1 B_u^-$ state for $10 \le N \le 22$.

Recent theoretical studies using highly accurate ab initio methods, however, suggest that internal conversion in carotenoids from the $1^1 B_u^+$ state to the $2^1 A_g^-$ state can proceed directly, without involving intermediate dark states.\cite{Taffet2019e,Khokhlov2020b} In order to investigate this particular mechanism, we choose a second set of $U$ and $V$ parameters, namely $U=7.25$ eV and $V=3.25$ eV. The vertical and relaxed excitation energies for this parameterisation is illustrated in Figure 1 of the Supporting Information. For this parameter regime, we find that for all relevant chain lengths the $2^1 A_g^-$ relaxed energy is lower than the $1^1 B_u^+$ relaxed energy, and that the $2^1 A_g^-$ vertical energy is higher than the $1^1 B_u^+$ vertical energy. Therefore, direct internal conversion from $1^1 B_u^+$ state to $2^1 A_g^-$ state is energetically possible for this parameter set.

Since $\hat{H}_{\textrm{UVP}}$ is invariant to particle-hole exchange, a symmetry breaking term $\hat{H}_\epsilon$ is introduced into our model to facilitate internal conversion from the $1^1 B_u^+$ state, which has positive particle-hole symmetry, to the triplet-pair states, with have negative particle-hole symmetry. This term is given by
\begin{equation}
	\hat{H}_\epsilon
		= \sum_{n=1}^{N} \epsilon_n (\hat{N}_n -1), \label{eqn:HSB}
\end{equation}
where $\epsilon_n$ is the  potential energy  on the $n^{\textrm{th}}$ C-atom.

We denote the $i^{\textrm{th}}$ singlet eigenstate of the full Hamiltonian $\hat{H}=(\hat{H}_{\textrm{UVP}}+\hat{H}_\lambda+\hat{H}_\epsilon)$ as $S_i$ (where $S_0$ is the ground state). We define these to be adiabatic states. We note that for $V=2.75$ eV (as in ref \citen{Manawadu2022}), $S_2$ is taken to be the initial state, $\Psi(t=0)$, at time $t=0$, as it has the largest $1^1 B_u^+$ character. Similarly, for $V=3.25$ eV, $\Psi(t=0) =S_1$, as it has the largest $1^1 B_u^+$ character.

As described in ref \citen{Manawadu2022a}, $\hat{H}_\epsilon$ is optimised under the constraint $\epsilon_n < \epsilon_{\textrm{max}}$ such that the ground state $\pi$-electron density on the $n^{\textrm{th}}$ C-atom reproduces the Mulliken charge densities of the $\pi$-system found \emph{via} ab initio density functional theory  (DFT) calculations.
The optimized $\hat{H}_\epsilon$ is given in Table 1 in the Supporting Information. The cut-off $\epsilon_{\textrm{max}}=1.0$ eV is chosen such that $\Psi(t=0)$ retains sufficient $1^1 B_u^+$ character, while accurately reproducing the DFT densities.


The electronic states and the ground state equilibrium geometry are determined via the static DMRG method, while the time evolution of the initially prepared photoexcited singlet is determined via adaptive tDMRG.\cite{Daley2004a,White2004b}
The nuclear degrees of freedom are treated classically \emph{via} the Ehrenfest equations of motion.
The theoretical and computational techniques employed here are described in full detail in the accompanying methodology paper.\cite{Manawadu2022a}


\section{`Bright' to `dark' state internal conversion}

We begin our discussion of `bright' to `dark' state internal conversion by describing our dynamical simulations in section \ref{Se:3.1}. These simulations predict that interstate conversion does happen with 10 fs.  In section \ref{frenkel} we then explain how interstate conversion occurs and why it is so fast.

\subsection{Computational results}\label{Se:3.1}

All of our calculations are performed on the carotenoid neurosporene, whose chemical formula is illustrated in Figure 1.
\begin{figure}[h!]
\centering	
         \includegraphics[height=0.05\textheight]{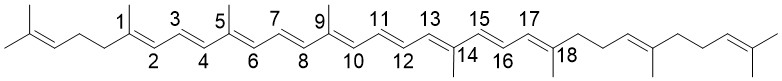}
           \caption{The structural formula of neurosporene, illustrating the 18 C-atom (9 double-bond) $\pi$-conjugated system.}
   \label{Figure2}
\end{figure}

\subsubsection{Internal conversion  from the $1^1 B_u^+$ to $2^1 A_g^-$ states via the $1^1 B_u^-$ state} \label{sec:1BuM}

As neurosporene does not possess $C_{2}$ symmetry, transitions from  $B_u$ to  $A_g$ states are not symmetry forbidden. This gives rise to a complex dynamical relaxation process involving the $1^1 B_u^+$, $1^1 B_u^-$, and $2^1 A_g^-$ diabatic states. This is in contrast to our previous work on zeathanxin\cite{Manawadu2022}, which does possess $C_{2}$ symmetry, and therefore only exhibits $1^1 B_u^+$ to $1^1 B_u^-$ state interconversion.

\begin{figure}[h!]
\centering
         \includegraphics{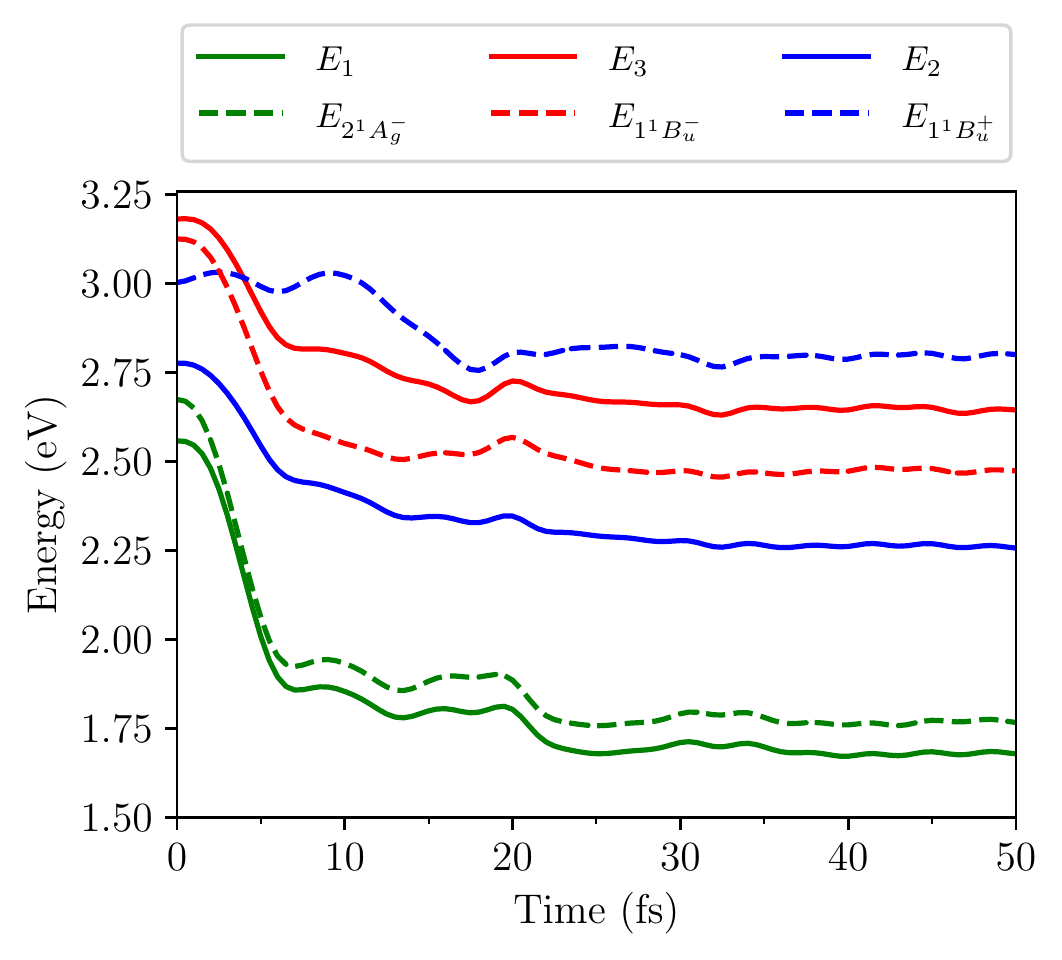}
  \caption{Excitation energies as a function of time of the diabatic $2^1 A_g^-$, $1^1 B_u^+$ and $1^1 B_u^-$ states, and the adiabatic $S_1$, $S_2$ and $S_3$ states. These energies are found for neurosporene  with $V=2.75$ eV. The $1^1 B_u^+$ and $1^1 B_u^-$ energies exhibit a  crossover at $\sim 5$ fs, while the $S_2$ and $S_3$ energies exhibit an avoided crossing.}
   \label{Figure3}
\end{figure}

For the parameter set $V = 2.75$ eV the initial photoexcited system $\Psi (t)$ is prepared in the adiabatic state $S_2$, as this has the largest overlap with the diabatic $1^1 B_u^+$ state at the Franck-Condon point. The nuclei begin in the ground state geometry and experience resultant forces exerted by the electrons, which causes $\hat{H}_{\textrm{e-n}}$ to change, and initiates the evolution of the electronic and nuclear degrees of freedom. Figure \ref{Figure3} shows the calculated adiabatic and diabatic excited energies as a function of time. A crossover of the diabatic  $1^1 B_u^-$ and $1^1 B_u^+$ energies occurs at $\sim 5$ fs, while the adiabatic $S_2$ and $S_3$ energies exhibit an avoided crossing, as the coupling between the diabatic states is nonzero, i.e., $\mel{1^1 B_u^+}{\hat{H}_\epsilon}{1^1 B_u^-} \neq 0$.

\begin{figure}[h!]
\centering
         \includegraphics{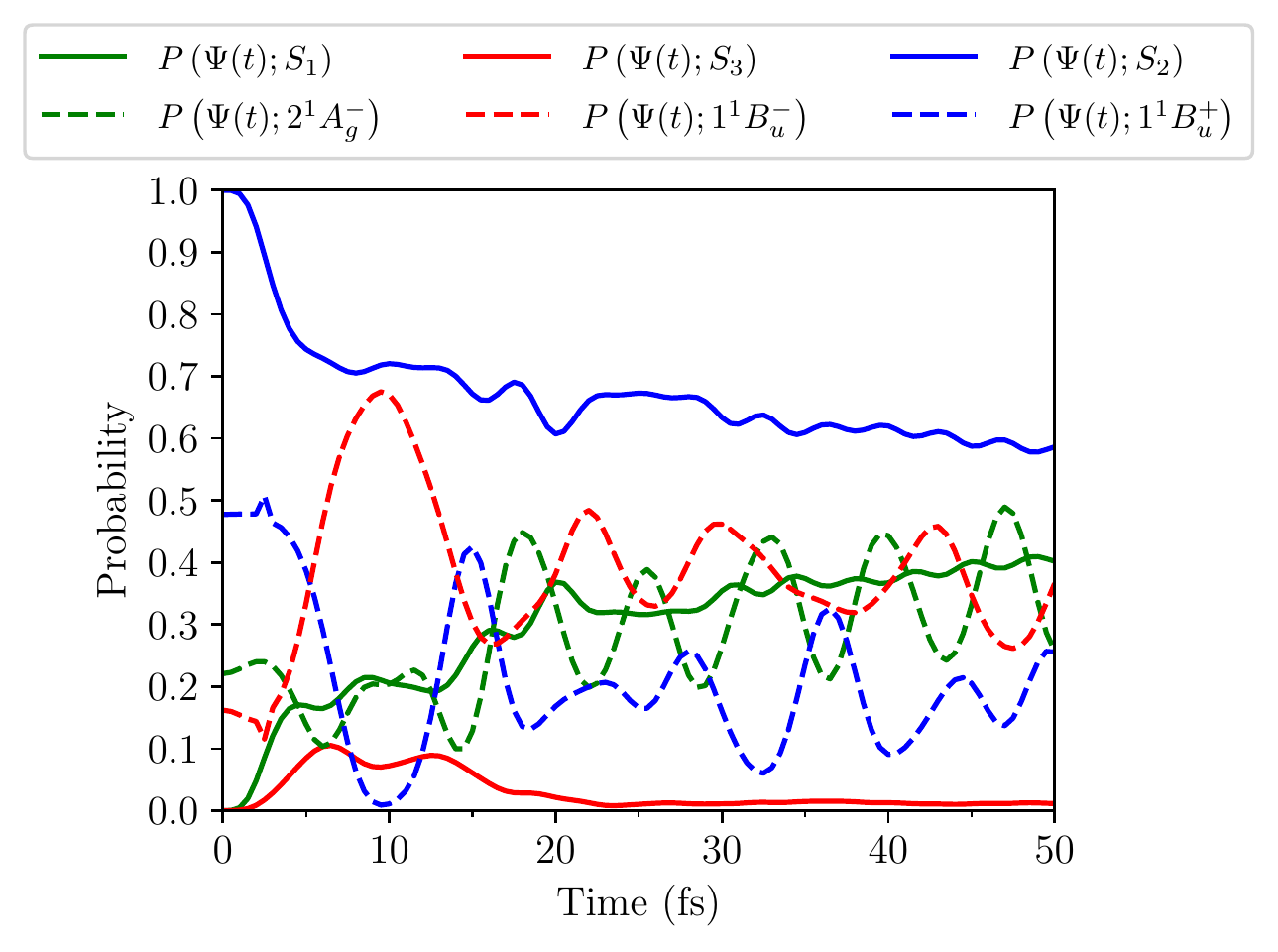}
  \caption{Probabilities as a function of time that the system described by $\Psi(t)$ occupies the adiabatic states, $S_1$, $S_2$ and $S_3$, and the diabatic states, $2^1 A_g^-$, $1^1 B_u^+$ and $1^1 B_u^-$. The results are for neurosporene with $V=2.75$ eV. Note that $\Psi(t=0) = S_2$.}
   \label{Figure4}
\end{figure}

The triplet-pair nature of the system at time $t$ is determined by calculating the probabilities that the system described by $\Psi (t)$ occupies the triplet-pair states, $1^1 B_u^-$ and $2^1 A_g^-$. Figure \ref{Figure4} shows the probabilities that  $\Psi (t)$  occupies the diabatic states $2^1 A_g^-$, $1^1 B_u^+$ and $1^1 B_u^-$, and the adiabatic states $S_1$, $S_2$ and $S_3$. At the avoided crossing $\Psi (t)$ exhibits an adiabatic transition from the $1^1 B_u^+$ state to the $1^1 B_u^-$ state.
Within $10$ fs  $\Psi (t)$ predominantly occupies the $1^1 B_u^-$ diabatic state.

After the ultrafast adiabatic transition to the $1^1 B_u^-$ state, the system continues to undergo a slow nonadiabatic transition to the $2^1 A_g^-$ state. This is a consequence of the nonzero  coupling between the $1^1 B_u^-$ and $2^1 A_g^-$ states, i.e., $\mel{1^1 B_u^-}{\hat{H}_\epsilon}{2^1 A_g^-} \neq 0$.

As both $2^1 A_g^-$ and $1^1 B_u^-$ are triplet-pair states, and noting that $P(S_3; \Psi(t)) \sim 0$ in the long-time limit, we extend the singlet triplet-pair yield calculation for a two level system\cite{Manawadu2022,Manawadu2022a} to calculate the total triplet-pair state probability, $P_{\mathrm{classical}}$, as
\begin{align}
	P_{\mathrm{classical}} =& P(S_1;2^1 A_g^-) \times P(\Psi(t);S_1) +
					P(S_2;2^1 A_g^-) \times P(\Psi(t);S_2) \notag \\
					+& P(S_1;1^1 B_u^-) \times P(\Psi(t);S_1) +
					P(S_2;1^1 B_u^-) \times P(\Psi(t);S_2).
\end{align}
After $\sim 50$ fs this yield is $\sim 70 \%$. (The probabilities, $P(S_i,\phi_j)$, that the adiabatic state, $S_i$ occupies the diabatic state $\phi_j$, are shown in Figure 2 of the Supporting Information.)

\subsubsection{Direct internal conversion from the $1^1 B_u^+$ to $2^1 A_g^-$ states}

\begin{figure}[h!]
\centering
         \includegraphics{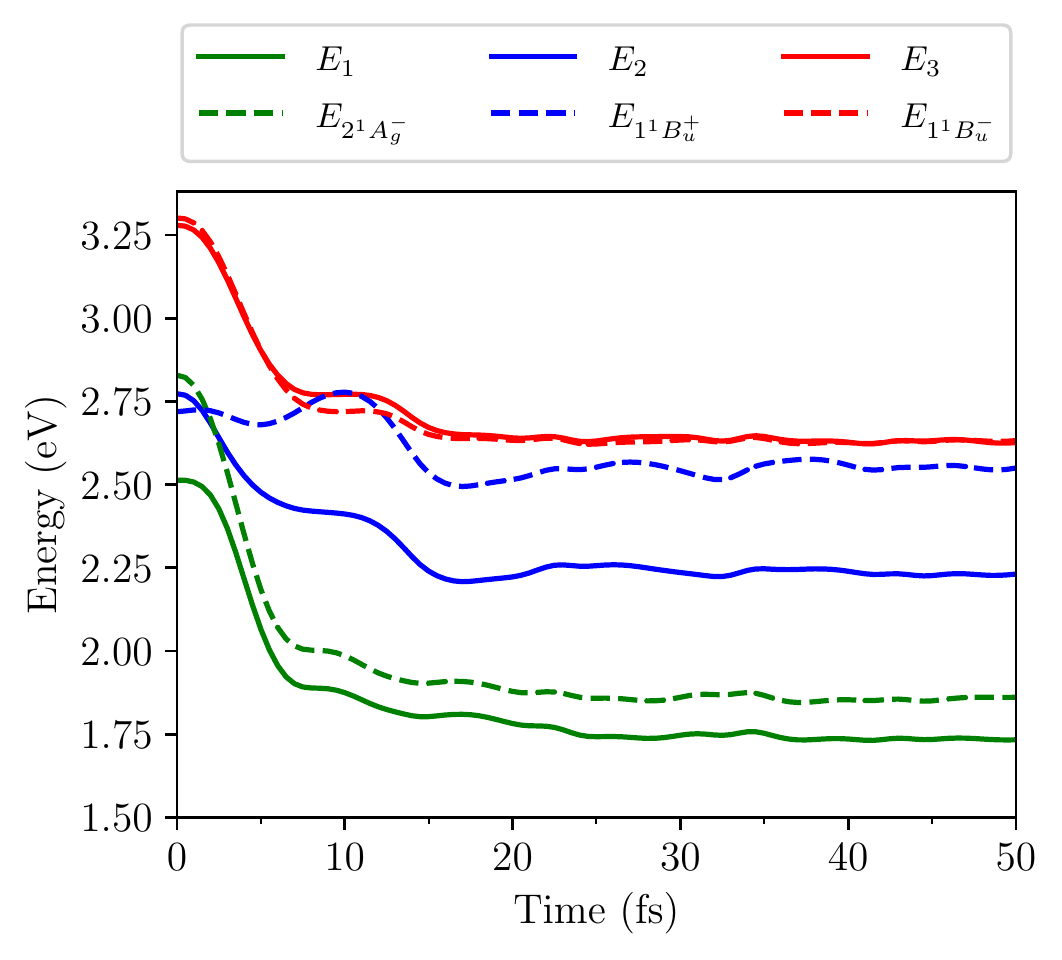}
  \caption{Excitation energies as a function of time of the diabatic $1^1 B_u^+$, $2^1 A_g^-$ and $1^1 B_u^-$ states, and the adiabatic $S_1$, $S_2$ and $S_3$ states. These energies are found for neurosporene with $V=3.25$ eV. The $1^1 B_u^+$ and $2^1 A_g^-$ energies exhibit a crossover at $\sim 3$ fs, while the $S_1$ and $S_2$ energies exhibit an avoided crossing.}
   \label{Figure6}
\end{figure}

As described in section 2, for the parameter set $V=3.25$ eV the primary photoexcited state is  $S_1$.
The singlet adiabatic and diabatic  energies  as a function of time are illustrated in Figure \ref{Figure6}. The   $2^1 A_g^-$ and $1^1 B_u^+$ energies  crossover within $\sim 3$ fs. In contrast, as a consequence of the diabatic coupling of the  $2^1 A_g^-$ and $1^1 B_u^+$ states,  the  $S_1$ and $S_2$ energies display an avoided crossing.

\begin{figure}[h!]
\centering
         \includegraphics{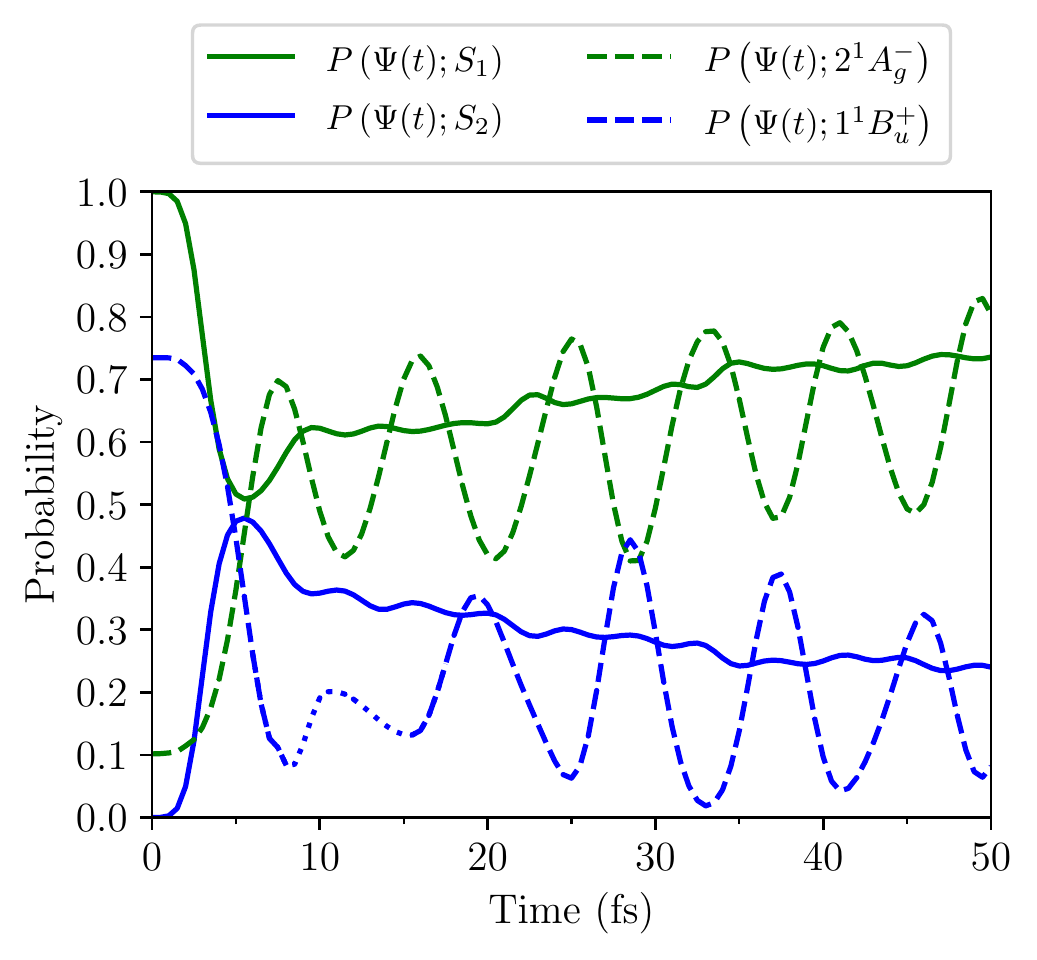}
  \caption{Probabilities as a function of time that the system described by $\Psi(t)$ occupies the adiabatic states, $S_1$ and $S_2$, and the diabatic states, $2^1 A_g^-$ and $1^1 B_u^+$. These results are for neurosporene  with $V=3.25$ eV.
  (Note that the dotted curve for the $1^1 B_u^+$ occupation is an interpolation of the computed data from 8 to 15 fs, as during this time the $1^1 B_u^+$ and $1^1 B_u^-$ energies are quasidegenerate, making it difficult to numerically resolve these wavefunctions.)
  }
   \label{Figure7}
\end{figure}

The probabilities that $\Psi(t)$  occupies  the  adiabatic and diabatic states, illustrated in Figure \ref{Figure7}, show that  the avoided crossing is accompanied by a transition of $\Psi(t)$ from the $1^1 B_u^+$ state to the $2^1 A_g^-$ state, while predominantly remaining in the $S_1$ state. The oscillations of the diabatic populations can be understood by a quasistationary two-state approximation, as discussed in  ref  \citen{Manawadu2022a}.

\begin{figure}[h!]
\centering
         \includegraphics{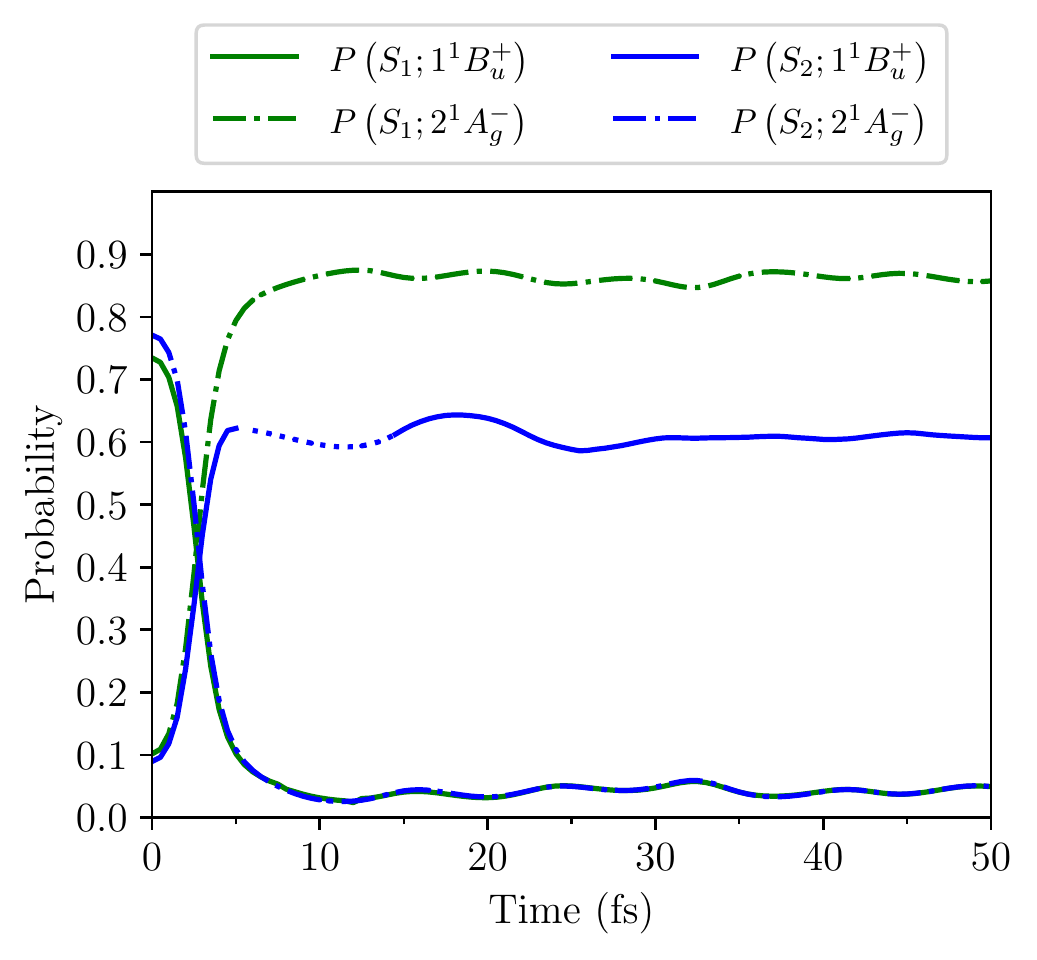}
  \caption{Probabilities as a function of time that the adiabatic states, $S_1$ and $S_2$, occupy the diabatic states, $2^1 A_g^-$ and $1^1 B_u^+$. These results are for neurosporene with $V=3.25$ eV.
    (Note that the dotted curve for the $1^1 B_u^+$ occupation is an interpolation of the computed data from 8 to 15 fs, as during this time the $1^1 B_u^+$ and $1^1 B_u^-$ energies are quasidegenerate, making it difficult to numerically resolve these wavefunctions.)
    }
   \label{Figure8}
\end{figure}

Further evidence for the adiabatic transition is provided by the calculated probabilities that the adiabatic states occupy the diabatic states, shown in Figure \ref{Figure8}. At $t=0$, the photoexcited state, $S_1$, primarily occupies the exciton state, $1^1 B_u^+$, while $S_2$ primarily occupies  the triplet-pair state, $2^1 A_g^-$. After the avoided crossing, at which the diabatic states contribute equally to the adiabatic states,  the $S_1$ state  predominantly occupies the $2^1 A_g^-$ state, while the $S_2$ state  predominantly occupies the $1^1 B_u^+$ state. The $2^1 A_g^-$ yield  after $\sim 50$ fs is $\sim 60 \%$.

\subsubsection{Transient spectra}

Transient (i.e., time-resolved) spectroscopy is an important experimental technique used in the study of carotenoid photophysics. Seminal work which utilised transient spectroscopy experiments include the measurement of the $S_1$ lifetime of carotenoids,\cite{Wasielewski1989} direct observation of the $S_1$ dark state,\cite{Polivka1999} and the detection of dark intermediate states between the $S_1$ and $S_2$ states.\cite{Gradinaru2001b,Fujii2003}

The theoretical transient absorption spectrum from state $S_i$ at time $t$ is given by the expression
\begin{equation}
	I_{\mu}(t,\omega)
		= \frac{1}{\pi} \sum_n \left | \mel{n}{\hat{\mu}}{S_i(t)} \right |^2
		 \delta(E_{i} + \omega - E_n), \label{eq:spectrum}
\end{equation}
where $E_i$ is the energy of state $S_i$. In this section, we present our results for transient spectra calculations using the Lanczos-DMRG method.\cite{PhysRevB.52.R9827,Kuhner1999a} The computational methodology is described in ref \citen{Manawadu2022a}.

\begin{figure}[h!]
\centering
         \includegraphics{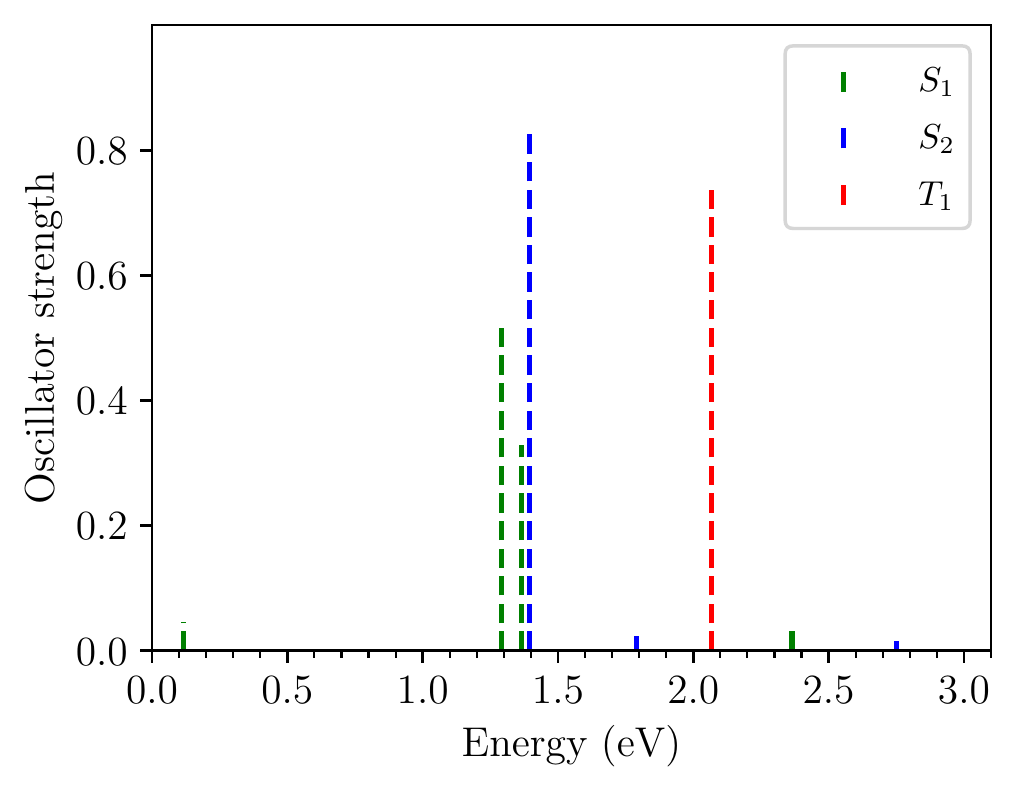}
         \caption{Calculated transient absorption spectra  at $t=0$ from the $S_1 \equiv 1^1 B_u^+$, $S_2 \equiv 2^1 A_g^-$ and $T_1$ states. Results are for a system with no broken-symmetry, i.e., $\hat{H}_\epsilon =0$.}
   \label{Figure9}
\end{figure}

We present results for $V=3.25$ eV, when there is direct $1^1 B_u^+$ to $2^1 A_g^-$ state conversion. First, we consider an ideal system with both particle-hole and $C_{2}$ symmetries (i.e., we set $\hat{H}_\epsilon =0$). Thus, at the Franck-Condon point (i.e., at $t=0$ fs) $S_1 \equiv 1^1 B_u^+$ and $S_2 \equiv 2^1 A_g^-$. The calculated initial absorption spectra from $S_1$ and $S_2$, and the triplet ground state, $T_1$, is shown in Figure \ref{Figure9}.
The three peaks arising from the $S_1$ state at $0.1$ eV, $1.3$ eV and $1.4$ eV are attributed to transitions from the $1^1 B_u^+$ state to the $2^1 A_g^-$, $4^1 A_g^-$ and $5^1 A_g^-$ states, respectively.

\begin{figure}[h!]
\centering
         \begin{subfigure}[b]{0.49\textwidth}
        		\includegraphics{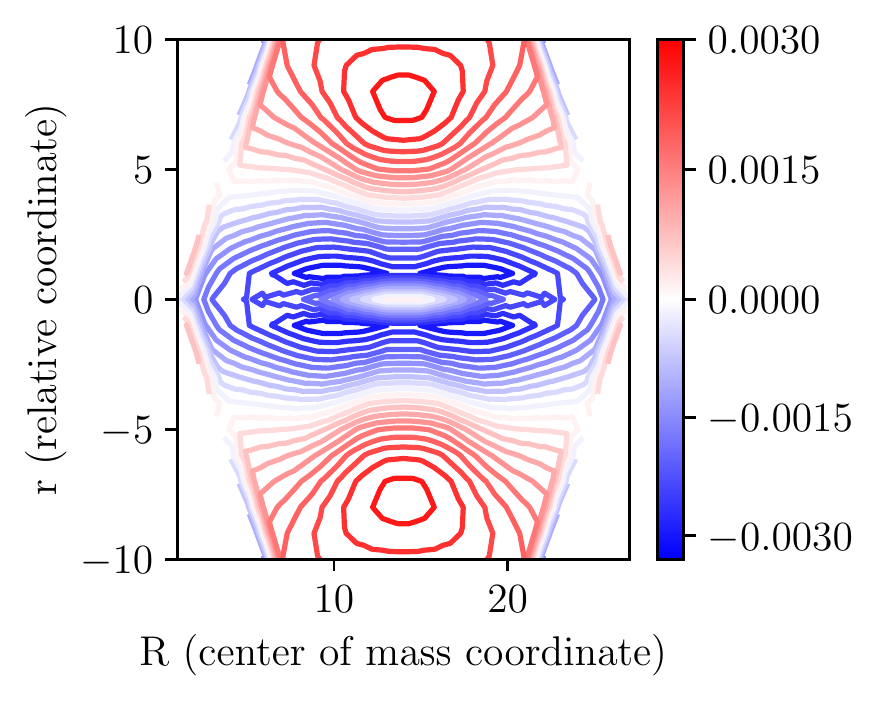}
         \end{subfigure}
         \caption{The $n^1 B_u^+$ exciton wave function calculated using eq 18 of ref \citenum{Valentine20}, using the PPP model with $N=54$.
         This has two nodal surfaces along the relative (i.e., electron-hole) coordinate and is thus dipole-connected to the charge-transfer exciton component of the $2^1 A_g^-$ state, which has one nodal surface along the relative coordinate.
         See ref \citenum{Valentine20} for the full parametrization. }
   \label{Figure10}
\end{figure}

The $2^1 A_g^-$ state is comprised of a singlet triplet-pair component and an odd-parity charge-transfer exciton component (see section \ref{frenkel}, and refs \citen{Valentine20} and \citen{Barford2022}). If the absorption signal from the $2^1 A_g^-$ state arose from a transition from its bound triplet-pair component to $T_1T_1^*$, we would  expect  this  to be higher in energy than  the $T_1$ to $T_1^*$  transition.\cite{Polak2019} We note, however, that this absorption peak is significantly lower in energy compared to the lowest absorption from the $T_1$ state. In contrast, if the absorption signal from the $2^1 A_g^-$ state arose from a transition from its charge-transfer exciton component, the resulting $n^1 B_u^+$ state would be an even-parity exciton.
We verify this latter hypothesis by calculating the exciton wave function of the  $n^1 B_u^+$ state; this is  shown in Figure \ref{Figure10}.\footnote{Readers are referred to ref \citen{Valentine20} for a calculation of exciton wave functions in linear polyenes and to ref \citen{Barford2013c} for a  discussion of excitons in conjugated polymers.} By observing the nodal structure of the $n^1 B_u^+$  exciton wave function, we conclude that the $2^1 A_g^-$ to $n^1 B_u^+$ transition does indeed arise from the charge-transfer exciton component of the $2^1 A_g^-$ state to  a Mott-Wannier exciton\cite{Barford2013c}.

Next, we consider the realistic Hamiltonian with the inclusion of the symmetry breaking term, $\hat{H}_\epsilon$.  For this parameter set the  $1^1 B_u^+$ and $2^1 A_g^-$ energies exhibit a crossover,  while the  $S_1$ and $S_2$ energies exhibit an avoided crossing (as shown in Figure \ref{Figure6}).

\begin{figure}[h!]
\centering
         \includegraphics{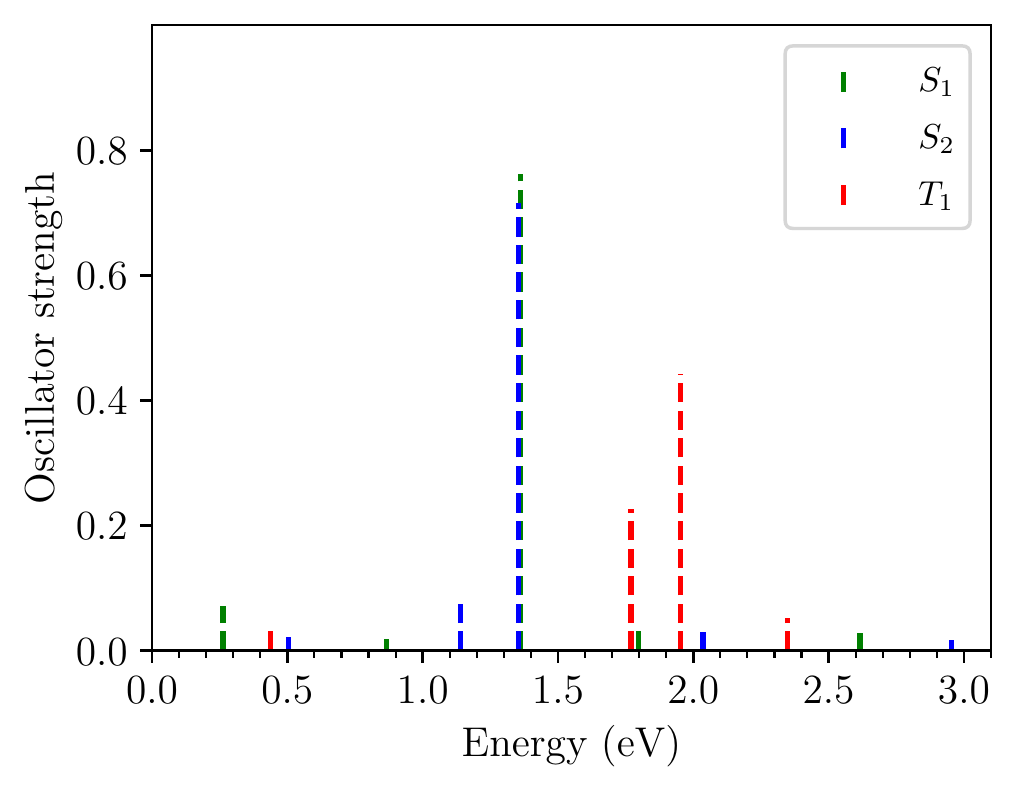}
         \caption{Calculated transient absorption spectra  at $t=0$ from the $S_1$, $S_2$ and $T_1$ states of neurosporene at $t=0$ fs. Note the overlapping $S_1$ and $S_2$ absorption at $\sim 1.4$ eV.}

   \label{Figure11}
\end{figure}

The calculated absorption spectra at the Franck-Condon point from the    $S_1$ and $S_2$ states, and the triplet ground state, $T_1$, is shown in Figure \ref{Figure11}.
At $t=0$ fs, $S_1$  predominantly occupies  the $1^1 B_u^+$ state. The absorption at $\sim 0.3$ eV corresponds to the $S_1 \rightarrow S_2$ transition, while the absorption maxima observed around $1.4$ eV is the transition from the $1^1 B_u^+$ state to a high energy $^1 A_g^-$ state.
%
%
Due to the inclusion of the symmetry breaking term, the adiabatic state $T_1$ will have some diabatic triplet excited state character, and therefore two absorption peaks are observed from $T_1$. As for Figure \ref{Figure9}, the transition from the charge-transfer exciton component of the $2^1 A_g^-$ state to the $n^1 B_u^+$ state at $\sim 1.4$ eV is lower in energy than the triplet absorption peaks.

\begin{figure}[h!]
\centering
	        \includegraphics{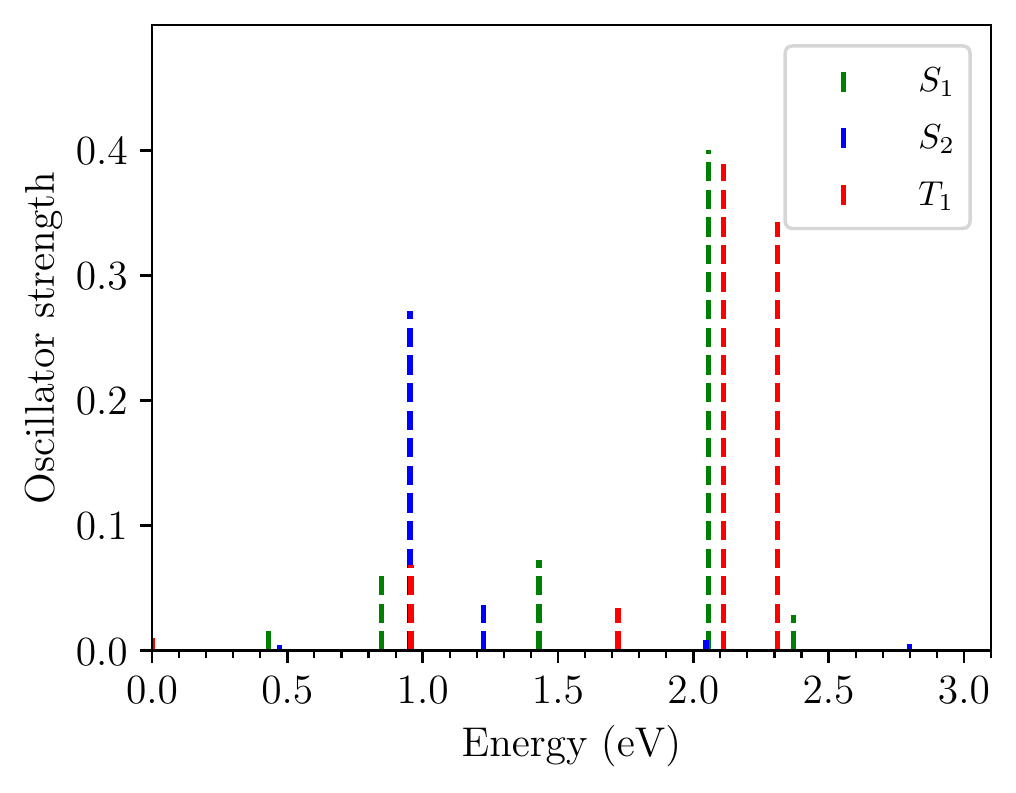}
  \caption{Calculated transient absorption spectra  at $t=20$ fs from the $S_1$, $S_2$ and $T_1$ states of neurosporene. The singlet state absorption spectra are weighted by $P(\Psi(t);S_i)$.}
   \label{Figure12}
\end{figure}

As the evolving photoexcited system, $\Psi(t)$, reaches equilibrium it  occupies  both the $S_1$ and $S_2$ states. Figure \ref{Figure12} shows the weighted absorption spectra at $t=20$ fs of these states, as well as the absorption spectra of the $T_1$ state. After passing through the avoided crossing at $\sim 3$ fs the $S_1$ state now predominantly occupies the $2^1 A_g^-$ state, while the $S_2$ state  predominantly  occupies the  $1^1 B_u^+$ state. This change of occupations is reflected in the calculated transient spectra, as the major signals originating from the $S_1$ state is now higher in energy compared to those originating from the $S_2$ state. At the relaxed geometry, the triplet peaks are blue shifted to $2.1$ eV and $2.3$ eV, which are close to the relaxed excited state values previously calculated for polyenes.\cite{Valentine20}

As the  $S_1$ state at 20 fs now predominantly occupies the $2^1 A_g^-$  state, we attribute the absorption peak of the $S_1$ state at $\sim 2.1$ eV to a transition from the charge-transfer exciton component of the $2^1 A_g^-$ state. This energy is blue-shifted from the corresponding transition from the $2^1 A_g^-$ component  of the $S_2$ state at $t=0$, because of the large reorganization energy of the $2^1 A_g^-$ state.
This energy is close to the experimentally observed $2^1 A_g^- \rightarrow n^1 B_u^+$ transition of carotenoids.\cite{Wasielewski1989,Polivka2009,Polak2019}
We also note that the energy of the major absorption from  $S_2$ around $0.9$ eV agrees with the experimentally observed $1^1 B_u^+ \rightarrow m^1 A_g^-$ transition.\cite{Zhang2001a}

In summary, our transient absorption calculations predict the following. At the Franck-Condon point there will be a photoexcited absorption at ca.\ 1.4 eV resulting from the  $1^1 B_u^+$ to $ n^1 A_g^-$ transition (and a weak transition at ca.\ 0.3 eV resulting from the  $1^1 B_u^+$ to $2^1 A_g^-$ transition). Within 20 fs, however, the adiabatic evolution of $S_1$ from $1^1 B_u^+$ to $2^1 A_g^-$ character results in a new transition from the charge-transfer exciton component of the $2^1 A_g^-$ state at ca.\ 2.0 eV, while a weaker transition at ca.\ 0.9 eV arises from the residual $1^1 B_u^+$ component of the evolving wavefunction.

\subsection{How and why does `bright' to `dark'  state internal conversion occur?} \label{frenkel}

As we have seen in the previous two sections, as a consequence of diabatic energy-level crossings, internal conversion from the optically `bright' state  to the `dark' state occurs within 10 fs. In this section we address the questions of how and why this process occurs.

As described in ref \citen{Barford2022}, the $2^1A_g^-$ state is a linear combination of a singlet triplet-pair and an odd-parity charge-transfer exciton.
Triplet-pair binding occurs because when a pair of triplets occupy neighboring ethylene dimers a one-electron transfer converts them to the odd-parity charge-transfer exciton. This hybridization causes a nearest neighbor triplet attraction. Similarly, the $1^1B_u^+$ state is predominately a Frenkel exciton (i.e., an electron-hole bound on the same dimer). However, the $1^1B_u^+$ state also consists of some even-parity charge-transfer exciton components.\footnote{The even(odd)-parity is indicated by the positive(negative) particle-hole symmetry label on the term symbol.}

\begin{figure}
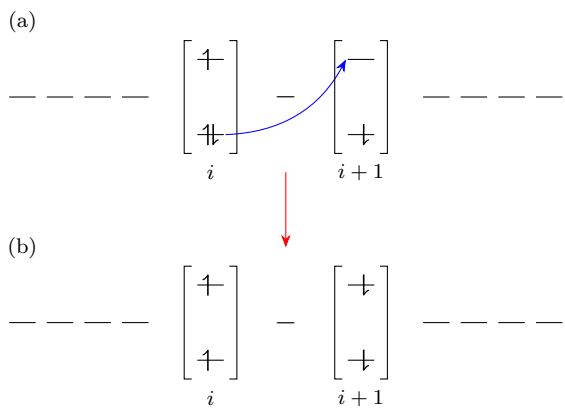

\begin{modiagram}
		\AO{s}{4;s=0} \AO(0.5cm){s}{4;s=0} \AO(1.0cm){s}{4;s=0} \AO(1.5cm){s}{4;s=0}
		\AO(2.5cm){s}{3.5}
		\AO(2.5cm){s}{4.5;up}
		 \node[align=center,font=\small] at (3.5cm,4) {$-$};
		 \AO(4.5cm){s}{3.5;down}
		\AO(4.5cm){s}{4.5;s=0}
		\AO(5.5cm){s}{4;s=0} \AO(6.0cm){s}{4;s=0} \AO(6.5cm){s}{4;s=0} \AO(7cm){s}{4;s=0}

		\node[align=center,font=\scriptsize] at (2.5,3) {$i$};
		\node[align=center,font=\scriptsize] at (4.5,3) {$i+1$};
		\draw [blue] (2.7,3.5) edge[bend right, -{Stealth[]}] (4.3,4.5) ;

		 \draw [-{Stealth},red] (3.5,3)--(3.5,2);

		\AO{s}{1;s=0} \AO(0.5cm){s}{1;s=0} \AO(1.0cm){s}{1;s=0} \AO(1.5cm){s}{1;s=0}
		\AO(2.5cm){s}{0.5;up}
		\AO(2.5cm){s}{1.5;up}
		 \node[align=center,font=\small] at (3.5cm,1) {$-$};
		 \AO(4.5cm){s}{0.5;down}
		\AO(4.5cm){s}{1.5;down}
		\AO(5.5cm){s}{1;s=0} \AO(6.0cm){s}{1;s=0} \AO(6.5cm){s}{1;s=0} \AO(7cm){s}{1;s=0}
		
		\node[align=center,font=\scriptsize] at (2.5,0) {$i$};
		\node[align=center,font=\scriptsize] at (4.5,0) {$i+1$};
		
		\draw [black] (2.25,3.25) to [square left brace ] (2.25,4.75);
		\draw [black] (2.75,3.25) to [square right brace ] (2.75,4.75);
		
		\draw [black] (4.25,3.25) to [square left brace ] (4.25,4.75);
		\draw [black] (4.75,3.25) to [square right brace ] (4.75,4.75);
		
		\draw [black] (2.25,0.25) to [square left brace ] (2.25,1.75);
		\draw [black] (2.75,0.25) to [square right brace ] (2.75,1.75);
		
		\draw [black] (4.25,0.25) to [square left brace ] (4.25,1.75);
		\draw [black] (4.75,0.25) to [square right brace ] (4.75,1.75);
		
		\node[align=center,font=\scriptsize] at (0,5) {(a)};
		\node[align=center,font=\scriptsize] at (0,2) {(b)};

		\end{modiagram}
\caption{A schematic diagram illustrating the internal conversion of the odd-parity charge-transfer component of the $1^1B_u$ state (a) to the nearest neighbor singlet triplet-pair component of the $2^1A_g$ state (b). Note that (a) is not a spin-symmetrized state; see ref \citenum{Barford2022} for an illustration of correctly spin-symmetrized singlet charge-transfer states.}
\label{Figure13}
\end{figure}

In practice, because of substituent side-groups, carotenoids do not possess definite particle-hole symmetry, and so neither do their electronic states. This means that the $2^1A_g$ state possesses some charge-transfer exciton components of \emph{even}-parity, while the $1^1B_u$ state possesses some charge-transfer exciton components of \emph{odd}-parity. As illustrated in Figure \ref{Figure13}, it is these odd-parity charge-transfer components of the $1^1B_u$ state which readily interconvert to singlet triplet-pairs, and thus cause the `bright' to `dark' state internal conversion. In spirit, this is the same mechanism proposed to explain singlet triplet-pair production in acene dimers\cite{Smith2010}, the difference being that acene molecules replace the ethylene dimers of carotenoid chains.

Evidently, the greater the deviation from perfect particle-hole symmetry, the greater the amount of covalent and ionic mixing in the electronic eigenstates. This then implies a larger coupling and a higher internal conversion yield between the bright and dark states. In our model, the deviation from particle-hole symmetry is represented by $\hat{H}_\epsilon$ (eq (\ref{eqn:HSB})) and in particular by the value of $\{\epsilon_n \}$. In our companion paper, ref \citen{Manawadu2022a}, we quantify this statement by computing the probabilities of adiabatic and diabatic transitions as a function of $\{\epsilon_n \}$. As $\epsilon_n \rightarrow 0$ the probability of a $1^1B_u^+$ to $2^1A_g^-$ transition vanishes, while the probability of a $S_1$ to $S_2$ transition becomes unity. Conversely, for large $\epsilon_n$ the probability of a $S_1$ to $S_2$ transition vanishes and the probability of a $1^1B_u^+$ to $2^1A_g^-$ transition increases.

Next, we explain \emph{why} internal conversion happens within 10 fs, i.e., we address the equivalent question of why after photoexcitation there is a diabatic energy-level crossing of the Frenkel exciton and singlet triplet-pair state. The answer to this question lies at the heart of what makes the electronic properties of linear polyenes so fascinating. It involves the roles of both electron-electron interactions and electron-nuclear coupling. We approach this question in three ways.

We first consider the electronic states of linear polyenes in the absence of electron-nuclear coupling, i.e., when electronic interactions dominate. In this case polyenes are Mott-Hubbard insulators: there is a charge (correlation) gap between the ground state (i.e., $1^1A_g^-$) and the lowest-lying ionic state (i.e., $1^1B_u^+$)\cite{Essler2005,Barford2013c}. The ground state is  a quantum Heisenberg antiferromagnet with, in the limit of infinitely long chains, a gapless spin density wave or triplet excitation (i.e.,  the $1^3B_u^-$ state). These triplets weakly bind to form gapless singlet triplet-pairs (i.e., the $2^1A_g^-$ state). Thus, in this limit there is a large spin-correlation gap between the $1^3B_u^-$ and  $1^1B_u^+$ states, and the `dark' state lies energetically below the `bright' state.

Next, consider  electronic states in the absence of electronic interactions, but when electron-nuclear coupling dominates. In this case polyenes are Peierls insulators, i.e., there is a gap between the filled valence band and the empty conduction band as a result of the incipient bond order wave causing bond dimerization\cite{Barford2013c}. Now the $1^3B_u^-$ and  $1^1B_u^+$ states both have an excitation gap and are degenerate. The  $2^1A_g^-$ state lies higher in energy. As already stated, the ground state is dimerized. The $1^3B_u^-$ and  $1^1B_u^+$ states, however, exhibit solitonic structures and a reversal of the bond dimerization in the middle of the chain from the ground state dimerization. Crucially, the solitons of the triplet ($1^3B_u^-$ state) are associated with spin-1/2 particles, i.e., spin radicals or spinons, while  solitons of the singlet ($1^1B_u^+$ state) are associated with an electron or hole.\cite{Barford2013c}

Finally, we consider the intermediate case, relevant for polyenes, when  electronic interactions and electron-nuclear coupling are both important. Now, the ground state dimerization is enhanced\cite{Dixit1984,Barford2013c}. The $1^3B_u^-$ and  $1^1B_u^+$ states exhibit a large spin-correlation gap. As a consequence, the  $2^1A_g^-$ state has significant triplet-pair character, causing  the  $1^1B_u^+$ and $2^1A_g^-$ states to be quasidegenerate in the ground state geometry (i.e., at the Franck-Condon point)\cite{Barford2022}. The relaxed geometries of the $1^3B_u^-$ and  $1^1B_u^+$ states are now quite different. The relaxed geometry of the $1^3B_u^-$ state is similar to the non-interacting limit: there are soliton-antisoliton structures associated with the spin-radicals and a reversal of bond dimerization from the ground state. However, the relaxed $1^1B_u^+$ geometry is quite different from the non-interacting limit: as a consequence of the electron-hole interaction, the soliton and antisoliton attract forming an exciton-polaron whose bond dimerization is only slightly different from the ground state. Thus, the $1^3B_u^-$ state exhibits a large reorganization energy, while the   $1^1B_u^+$ state does not. Similarly, the $2^1A_g^-$ state, being composed of a triplet-pair, also exhibits a large reorganization energy.

In summary, the reasons that there is an energy level crossing between the diabatic bright and dark states are the following. First, because of the large spin-correlation gap, the dark state has a large triplet-pair component and thus the bright and dark states are quasidegenerate at the Franck-Condon point. Second, the triplet state ($1^3 B_u^-$) has a larger reorganization energy than the $1^1 B_u^+$ state and consequently, because of its triplet-pair character, the reorganization of the dark state is much larger than for the bright state. Consequently, after photoexcitation to the bright state, nuclear forces cause the level crossing, and hence ultrafast internal conversion to the dark state.


\section{Exothermic intermolecular singlet fission}\label{}

As shown in Figure 1 of the Supporting Information, and refs \citen{Valentine20} and \citen{Manawadu2022}, the intramolecular triplet-pair binding energy varies from ca.\ $1$ eV in short carotenoids to ca.\ $0.3$ eV in long polyene chains. Thus, intramolecular singlet fission is a strongly endothermic process, consistent with the absence of experimental evidence for free triplets on isolated carotenoids generated via a singlet fission mechanism\cite{Musser2015}.

In this section we address the question of how  intermolecular singlet fission can in principle be an exothermic process. Clearly, triplets on single molecules must be energetically stabilized to overcome the intramolecular  triplet-pair binding. There are two causes for this. First, there is quantum deconfinement: a single triplet delocalized on a whole molecule has a smaller kinetic (or zero point) energy than a single triplet confined to half a molecule, which is the relevant comparison for two unbound triplets on the same chain. That is, it costs less energy to unbind a pair of triplets on separate molecules than on the same molecule. Second, self-localized triplets on two separate molecules experience a larger (negative) reorganization energy than two bound triplets on the same molecule. Crucially, there are two components to the reorganization energy:  a term arising from C-C bond stretches and an additional term arising from torsional relaxation. Therefore, for exothermic intermolecular singlet fission to occur, the molecules must exist in an environment so that they have twisted  ground state conformations.

To quantify these statements, we supplement the UVP model (introduced in section \ref{sec:comp}) by terms that model electron-torsional coupling. The $\pi$-electron transfer integral is $\beta(\theta) = \beta_0 \cos \theta$, where $\theta$ is the dihedral angle between neighboring C-H groups. Assuming a small planarization in the excited state, i.e., assuming that $\delta \theta \ll \theta^0$, we may write
\begin{equation}\label{}
  \beta(\theta) \approx \beta_0\left(\cos \theta^0 - \delta \theta\sin \theta^0 \right),
\end{equation}
where $\theta^0$ is the ground state equilibrium dihedral angle. Thus, $  - \beta_0\sin \theta^0$ is the electron-torsional coupling parameter that couples the bond-order operator, $\hat{T}$, to the variation in dihedral angle, $\delta \theta$. We also assume that there is a harmonic elastic energy
\begin{equation}\label{}
  E_{rot} = \frac{1}{2} K_{rot} \delta \theta^2.
\end{equation}

Applying the Hellmann-Feynman theorem implies that the equilibrium dihedral angle for the $n^\textrm{th}$ bond is $\theta_n = \theta^0_n + \delta \theta_n$, where
\begin{equation}\label{}
 \delta \theta_n = - \frac{2 \langle T_n \rangle}{K_{rot}} \left( \beta_0 - 2\alpha \delta u_n \right) \sin \theta_n^0
\end{equation}
and $\delta u_n = (u_{n+1}-u_n)$ is the change of bond length caused by electronic coupling to C-C bond vibrations.

We define the intermolecular triplet-pair binding energy, $\Delta E_{TT}$, as twice the energy of triplets on separate molecules relative to the intramolecular triplet-pair state (i.e., $2A_g$), namely
\begin{equation}\label{Eq:101}
  \Delta E_{TT} = 2\times E_{T_1}(N) - E_{2A_g}(N),
\end{equation}
where $N$ is the number of conjugated carbon-atoms in each molecule. A negative value of $\Delta E_{TT}$ implies exothermic intermolecular singlet fission.

Our results are presented in Tables \ref{ta:1} - \ref{ta:3}. Table \ref{ta:1} shows the results for a pair of carotenoid chains  of 22 conjugated carbon-atoms each, for different ground state twists, $\theta^0$, and torsional force constants, $K_{rot}$. Evidently, in the absence of torsional relaxation (e.g., if $\theta^0 = 0$ or $K_{rot} \rightarrow \infty$) intermolecular singlet fission is endothermic. For a fixed $K_{rot}$ singlet fission becomes more exothermic for a more twisted molecule in the ground state. Similarly, for a fixed  ground state twist  singlet fission becomes more exothermic as $K_{rot}$ is reduced.

\begin{table}
\small\centering
{\renewcommand{\arraystretch}{1.2}
\begin{tabular}{|p{0.30\linewidth}|p{0.10\linewidth}p{0.10\linewidth}p{0.10\linewidth}p{0.10\linewidth}|}
\hline
\multirow{2}{*}{$\theta^0$ (degrees)}  & \multicolumn{4}{|c|}{$K_{rot}$ (eV rad$^{-2}$)}  \\
\cline{2-5}
  & 6 & 8 & 10 & $\infty$ \\
\hline
0 & $+102$ & $+102$ & $+102$ & $+102$   \\
5 & $+75$ & $+88$ & $+93$ & $+102$   \\
10 & $-5$ & $+46$ & $+68$ & $+102$   \\
15 & $-134$ & $-22$ & $+27$ & $+102$  \\
20 & --- & $-111$ & $-27$ & $+102$  \\
\hline
\end{tabular}}
\caption{Triplet-pair binding energies in meV, $\Delta E_{TT}$, defined in eq (\ref{Eq:101}), for a pair of carotenoid chains, both of 22 conjugated carbon-atoms. A negative value implies exothermic intermolecular singlet fission.}
\label{ta:1}
\end{table}

Table \ref{ta:2} indicates that for the same values of $\theta^0$ and $K_{rot}$ intermolecular singlet fission becomes less favourable as the number of conjugated carbon atoms increases. This is because the energy gained by quantum deconfinement reduces with increasing chain length.

\begin{table}
\small\centering
{\renewcommand{\arraystretch}{1.2}
\begin{tabular}{|p{0.20\linewidth}|p{0.30\linewidth}|}
\hline
$N$ & Binding energy (meV)  \\
\hline
18 & $-54$   \\
22 & $-22$  \\
26 & +2   \\
\hline
\end{tabular}}
\caption{Triplet-pair binding energies in meV for a pair  of carotenoid chains of $N$ conjugated carbon-atoms. $\theta^0 = 15^0$ and $K_{rot} = 8$ eV rad$^{-2}$.}
\label{ta:2}
\end{table}

Finally, Table \ref{ta:3} list the various contributions that favor exothermic intermolecular singlet fission. Quantum deconfinement onto two molecules reduces the intramolecular binding energy on 22-site chains from $781$ meV to $226$ meV; bond relaxation causes  a $124$ meV decrease in binding energy; while additional torsional relaxation  causes another  $124$ meV decrease, rendering the process exothermic. Evidently, quantum deconfinement causes the largest reduction in binding energy, but all three components are necessary to enable exothermic singlet fission. In particular, it is necessary that the molecules are twisted in their ground states.

\begin{table}
\small\centering
{\renewcommand{\arraystretch}{1.2}
\begin{tabular}{|p{0.22\linewidth}|p{0.22\linewidth}|p{0.22\linewidth}|p{0.22\linewidth}|}
\hline
Intramolecular vertical&Intermolecular vertical & Bond relaxation & Bond and torsional relaxation  \\
\hline
781 & 226 & 102 & $-22$   \\
\hline
\end{tabular}}
\caption{Triplet-pair binding energies in meV for a pair of carotenoid chains of $22$ conjugated carbon-atoms. $\theta^0 = 15^0$ and $K_{rot}=8$ eV rad$^{-2}$.}
\label{ta:3}
\end{table}

\section{Discussion and conclusions}

This paper has described our dynamical simulations of the excited states of the carotenoid, neurosporene, following its photoexcitation into the `bright' (nominally $1^1B_u^+$) state. We employed the adaptive tDMRG method on the UV model of $\pi$-conjugated electrons and used the Ehrenfest equations of motion to simulate the coupled nuclei dynamics.

To account for the experimental and theoretical uncertainty in the relative energetic ordering of the nominal $1^1B_u^+$ and $2^1A_g^-$  states at the Franck-Condon point, we considered two sets of parameters. In both cases there is ultrafast internal conversion from the `bright' state to a `dark' singlet triplet-pair state, i.e., to one member of the `$2A_g$' family of states.

For one parameter set, internal conversion from the $1^1B_u^+$ to $2^1A_g^-$ states occurs via the dark, intermediate $1^1B_u^-$ state. In this case there is a crossover of the $1^1B_u^+$ and $1^1B_u^-$ diabatic energies within 5 fs and an associated avoided crossing of the $S_2$ and $S_3$ adiabatic energies. Following the adiabatic evolution of the $S_2$ state from predominately $1^1B_u^+$ character to predominately $1^1B_u^-$ character, there is a slower nonadiabatic transition from $S_2$ to $S_1$, accompanied by an increase in the population of the $2^1A_g^-$  state.

For the other parameter set the $2^1A_g^-$ energy lies higher than the $1^1B_u^+$ energy at the Franck-Condon point. In this case there is crossover of the  $2^1A_g^-$ and  $1^1B_u^+$ energies and an avoided crossing of the $S_1$ and $S_2$ energies, as the $S_1$ state evolves adiabatically from being of $1^1B_u^+$ character to $2^1A_g^-$ character.

We make a direct connection from our predictions to experimental observables by calculating the transient absorption. For the case of direct $1^1B_u^+$ to $2^1A_g^-$ internal conversion, we showed that the dominant transition at ca.\ 2 eV, being close to but lower in energy than the $T_1$ to $T_1^*$ transition, can be attributed to the  $2^1A_g^-$ component of $S_1$. Moreover, we show that it is the charge-transfer exciton component of the  $2^1A_g^-$ state that is responsible for this transition (to a higher-lying exciton state), and not its triplet-pair component. This transition is blue-shifted from the Franck-Condon point, because of the large reorganization energy of the $2^1A_g^-$ state.

We next discussed the microscopic mechanism of `bright' to `dark' state internal conversion, emphasising that this occurs via the exciton components of both states.

Finally, we described a mechanism whereby the strongly bound intrachain triplet-pairs of the `dark' state may undergo interchain exothermic dissociation. This mechanism relies on the possibility of the unbound interchain triplets being energetically stabilized by quantum deconfinement, and larger bond and torsional reorganization energies. We predict that this is only possible if the molecules are twisted in their ground states.

Irrespective of the ordering of the $1^1B_u^+$ and $2^1A_g^-$ states at photoexcitation, our simulations indicate that after 50 fs the yield of the `dark', predominately singlet triplet-pair states is ca.\ 65 \%. This implies, however, that the evolving state still has some `bright' (i.e., $1^1B_u^+$) component, which explains the weak emissive character of photoexcited carotenoids.

In an earlier paper we explained the origin of the intrachain triplet-pair binding\cite{Barford2022}, while in this paper we argue that exothermic interchain  triplet-pair dissociation is possible if it is accompanied by torsional relaxation. Work is now in progress to build a full model of singlet triplet-pair dissociation and spin decoherence  to understand the full kinetic process of singlet fission in carotenoid dimers. Future work will also investigate our model with fully quantized phonons. This will enable us to calculate the vibronic lineshape of the photoinduced absorption spectra, in particular allowing for a comparison of the $S_1$ and $T_1$ transient absorption\cite{Polak2019}. We will also investigate the validity of the Ehrenfest approximation for the nonadiabatic $S_2$ to $S_1$ transition, discussed in section \ref{sec:1BuM}.

\begin{acknowledgement}

We thank Jenny Clark for helpful discussions about the photophysics of carotenoid systems. D.M is grateful to the EPSRC Centre for Doctoral Training, Theory and Modelling in Chemical Sciences, under Grant No. EP/L015722/1 and Linacre College for a Carolyn and Franco Gianturco Scholarship and the Department of Chemistry, University of Oxford for financial support.  We acknowledge the use of University of Oxford Advanced Research Computing (ARC) facility for this work.

\end{acknowledgement}

\begin{suppinfo}
The \emph{Supporting Information} contains the following  sections: \emph{1.\ Parametrization of the UV-Peierls Hamiltonian}, \emph{2.\ Parametrization of the symmetry breaking Hamiltonian, $\hat{H}_\epsilon$} and \emph{3.\ Probabilities that the adiabatic states, $S_1$, $S_2$ and $S_3$ occupy the diabatic states $2^1 A_g^-$, $1^1 B_u^+$ and $1^1 B_u^-$ for case (a)}.
\end{suppinfo}

\newpage
\bibliography{D.Phil-Writing-Experimental_JPCA}

\providecommand{\latin}[1]{#1}
\makeatletter
\providecommand{\doi}
  {\begingroup\let\do\@makeother\dospecials
  \catcode`\{=1 \catcode`\}=2 \doi@aux}
\providecommand{\doi@aux}[1]{\endgroup\texttt{#1}}
\makeatother
\providecommand*\mcitethebibliography{\thebibliography}
\csname @ifundefined\endcsname{endmcitethebibliography}
  {\let\endmcitethebibliography\endthebibliography}{}
\begin{mcitethebibliography}{36}
\providecommand*\natexlab[1]{#1}
\providecommand*\mciteSetBstSublistMode[1]{}
\providecommand*\mciteSetBstMaxWidthForm[2]{}
\providecommand*\mciteBstWouldAddEndPuncttrue
  {\def\EndOfBibitem{\unskip.}}
\providecommand*\mciteBstWouldAddEndPunctfalse
  {\let\EndOfBibitem\relax}
\providecommand*\mciteSetBstMidEndSepPunct[3]{}
\providecommand*\mciteSetBstSublistLabelBeginEnd[3]{}
\providecommand*\EndOfBibitem{}
\mciteSetBstSublistMode{f}
\mciteSetBstMaxWidthForm{subitem}{(\alph{mcitesubitemcount})}
\mciteSetBstSublistLabelBeginEnd
  {\mcitemaxwidthsubitemform\space}
  {\relax}
  {\relax}

\bibitem[Fraser \latin{et~al.}(2001)Fraser, Hashimoto, and Cogdell]{Fraser2001}
Fraser,~N.~J.; Hashimoto,~H.; Cogdell,~R.~J. {Carotenoids and bacterial
  photosynthesis: The story so far...} \emph{Photosynthesis Research}
  \textbf{2001}, \emph{70}, 249--256\relax
\mciteBstWouldAddEndPuncttrue
\mciteSetBstMidEndSepPunct{\mcitedefaultmidpunct}
{\mcitedefaultendpunct}{\mcitedefaultseppunct}\relax
\EndOfBibitem
\bibitem[Uragami \latin{et~al.}(2020)Uragami, Sato, Yukihira, Fujiwara, Kosumi,
  Gardiner, Cogdell, and Hashimoto]{Uragami2020}
Uragami,~C.; Sato,~H.; Yukihira,~N.; Fujiwara,~M.; Kosumi,~D.; Gardiner,~A.~T.;
  Cogdell,~R.~J.; Hashimoto,~H. {Photoprotective mechanisms in the core LH1
  antenna pigment-protein complex from the purple photosynthetic bacterium,
  Rhodospirillum rubrum}. \emph{Journal of Photochemistry and Photobiology A:
  Chemistry} \textbf{2020}, \emph{400}, 112628\relax
\mciteBstWouldAddEndPuncttrue
\mciteSetBstMidEndSepPunct{\mcitedefaultmidpunct}
{\mcitedefaultendpunct}{\mcitedefaultseppunct}\relax
\EndOfBibitem
\bibitem[Hashimoto \latin{et~al.}(2016)Hashimoto, Uragami, and
  Cogdell]{Hashimoto2016}
Hashimoto,~H.; Uragami,~C.; Cogdell,~R.~J. {Carotenoids and Photosynthesis}.
  \emph{Subcellular Biochemistry} \textbf{2016}, \emph{79}, 111--139\relax
\mciteBstWouldAddEndPuncttrue
\mciteSetBstMidEndSepPunct{\mcitedefaultmidpunct}
{\mcitedefaultendpunct}{\mcitedefaultseppunct}\relax
\EndOfBibitem
\bibitem[Hudson and Kohler(1972)Hudson, and Kohler]{Hudson1972}
Hudson,~B.~S.; Kohler,~B.~E. {A low-lying weak transition in the polyene
  $\alpha$,$\omega$-diphenyloctatetraene}. \emph{Chemical Physics Letters}
  \textbf{1972}, \emph{14}, 299--304\relax
\mciteBstWouldAddEndPuncttrue
\mciteSetBstMidEndSepPunct{\mcitedefaultmidpunct}
{\mcitedefaultendpunct}{\mcitedefaultseppunct}\relax
\EndOfBibitem
\bibitem[Schulten and Karplus(1972)Schulten, and Karplus]{Schulten1972}
Schulten,~K.; Karplus,~M. {On the origin of a low-lying forbidden transition in
  polyenes and related molecules}. \emph{Chemical Physics Letters}
  \textbf{1972}, \emph{14}, 305--309\relax
\mciteBstWouldAddEndPuncttrue
\mciteSetBstMidEndSepPunct{\mcitedefaultmidpunct}
{\mcitedefaultendpunct}{\mcitedefaultseppunct}\relax
\EndOfBibitem
\bibitem[Hayden and Mele(1986)Hayden, and Mele]{Hayden86}
Hayden,~G.~W.; Mele,~E.~J. {Correlation-Effects and Excited-States in
  Conjugated Polymers}. \emph{Physical Review B} \textbf{1986}, \emph{34},
  5484\relax
\mciteBstWouldAddEndPuncttrue
\mciteSetBstMidEndSepPunct{\mcitedefaultmidpunct}
{\mcitedefaultendpunct}{\mcitedefaultseppunct}\relax
\EndOfBibitem
\bibitem[Tavan and Schulten(1987)Tavan, and Schulten]{Tavan1987}
Tavan,~P.; Schulten,~K. {Electronic excitations in finite and infinite
  polyenes}. \emph{Phys. Rev. B} \textbf{1987}, \emph{36}, 4337\relax
\mciteBstWouldAddEndPuncttrue
\mciteSetBstMidEndSepPunct{\mcitedefaultmidpunct}
{\mcitedefaultendpunct}{\mcitedefaultseppunct}\relax
\EndOfBibitem
\bibitem[Chandross \latin{et~al.}(1999)Chandross, Shimoi, and
  Mazumdar]{Chandross1999}
Chandross,~M.; Shimoi,~Y.; Mazumdar,~S. {Diagrammatic exciton-basis
  valence-bond theory of linear polyenes}. \emph{Phys. Rev. B} \textbf{1999},
  \emph{59}, 4822\relax
\mciteBstWouldAddEndPuncttrue
\mciteSetBstMidEndSepPunct{\mcitedefaultmidpunct}
{\mcitedefaultendpunct}{\mcitedefaultseppunct}\relax
\EndOfBibitem
\bibitem[Bursill and Barford(1999)Bursill, and Barford]{Bursill1999}
Bursill,~R.~J.; Barford,~W. {Electron-lattice relaxation, and soliton
  structures and their interactions in polyenes}. \emph{Physical Review
  Letters} \textbf{1999}, \emph{82}, 1514--1517\relax
\mciteBstWouldAddEndPuncttrue
\mciteSetBstMidEndSepPunct{\mcitedefaultmidpunct}
{\mcitedefaultendpunct}{\mcitedefaultseppunct}\relax
\EndOfBibitem
\bibitem[Barford \latin{et~al.}(2001)Barford, Bursill, and
  Lavrentiev]{Barford01}
Barford,~W.; Bursill,~R.~J.; Lavrentiev,~M.~Y. {Density-matrix
  renormalization-group calculations of excited states of linear polyenes}.
  \emph{Phys. Rev. B} \textbf{2001}, \emph{63}, 195108\relax
\mciteBstWouldAddEndPuncttrue
\mciteSetBstMidEndSepPunct{\mcitedefaultmidpunct}
{\mcitedefaultendpunct}{\mcitedefaultseppunct}\relax
\EndOfBibitem
\bibitem[Barford(2013)]{Barford2013c}
Barford,~W. \emph{{Electronic and optical properties of conjugated polymers}},
  2nd ed.; Oxford University Press: Oxford, 2013\relax
\mciteBstWouldAddEndPuncttrue
\mciteSetBstMidEndSepPunct{\mcitedefaultmidpunct}
{\mcitedefaultendpunct}{\mcitedefaultseppunct}\relax
\EndOfBibitem
\bibitem[Pol{\'{i}}vka and Sundstr{\"{o}}m(2009)Pol{\'{i}}vka, and
  Sundstr{\"{o}}m]{Polivka2009}
Pol{\'{i}}vka,~T.; Sundstr{\"{o}}m,~V. {Dark excited states of carotenoids:
  Consensus and controversy}. \emph{Chemical Physics Letters} \textbf{2009},
  \emph{477}, 1--11\relax
\mciteBstWouldAddEndPuncttrue
\mciteSetBstMidEndSepPunct{\mcitedefaultmidpunct}
{\mcitedefaultendpunct}{\mcitedefaultseppunct}\relax
\EndOfBibitem
\bibitem[Frank \latin{et~al.}(1997)Frank, Desamero, Chynwat, Gebhard, van~der
  Hoef, Jansen, Lugtenburg, Gosztola, and Wasielewski]{Frank1997}
Frank,~H.~A.; Desamero,~R. Z.~B.; Chynwat,~V.; Gebhard,~R.; van~der Hoef,~I.;
  Jansen,~F.~J.; Lugtenburg,~J.; Gosztola,~D.; Wasielewski,~M.~R.
  {Spectroscopic Properties of Spheroidene Analogs Having Different Extents of
  $\pi$-Electron Conjugation}. \emph{The Journal of Physical Chemistry A}
  \textbf{1997}, \emph{101}, 149--157\relax
\mciteBstWouldAddEndPuncttrue
\mciteSetBstMidEndSepPunct{\mcitedefaultmidpunct}
{\mcitedefaultendpunct}{\mcitedefaultseppunct}\relax
\EndOfBibitem
\bibitem[Kosumi \latin{et~al.}(2006)Kosumi, Yanagi, Fujii, Hashimoto, and
  Yoshizawa]{Kosumi2006}
Kosumi,~D.; Yanagi,~K.; Fujii,~R.; Hashimoto,~H.; Yoshizawa,~M. {Conjugation
  length dependence of relaxation kinetics in $\beta$-carotene homologs probed
  by femtosecond Kerr-gate fluorescence spectroscopy}. \emph{Chemical Physics
  Letters} \textbf{2006}, \emph{425}, 66--70\relax
\mciteBstWouldAddEndPuncttrue
\mciteSetBstMidEndSepPunct{\mcitedefaultmidpunct}
{\mcitedefaultendpunct}{\mcitedefaultseppunct}\relax
\EndOfBibitem
\bibitem[Santra \latin{et~al.}(2022)Santra, Ray, and Ghosh]{Santra2022}
Santra,~S.; Ray,~J.; Ghosh,~D. {Mechanism of Singlet Fission in Carotenoids
  from a Polyene Model System}. \emph{Journal of Physical Chemistry Letters}
  \textbf{2022}, \emph{13}, 6800--6805\relax
\mciteBstWouldAddEndPuncttrue
\mciteSetBstMidEndSepPunct{\mcitedefaultmidpunct}
{\mcitedefaultendpunct}{\mcitedefaultseppunct}\relax
\EndOfBibitem
\bibitem[Taffet \latin{et~al.}(2019)Taffet, Lee, Toa, Pace, Rumbles, Southall,
  Cogdell, and Scholes]{Taffet2019e}
Taffet,~E.~J.; Lee,~B.~G.; Toa,~Z. S.~D.; Pace,~N.; Rumbles,~G.; Southall,~J.;
  Cogdell,~R.~J.; Scholes,~G.~D. {Carotenoid Nuclear Reorganization and
  Interplay of Bright and Dark Excited States}. \emph{The Journal of Physical
  Chemistry B} \textbf{2019}, \emph{123}, 8628--8643\relax
\mciteBstWouldAddEndPuncttrue
\mciteSetBstMidEndSepPunct{\mcitedefaultmidpunct}
{\mcitedefaultendpunct}{\mcitedefaultseppunct}\relax
\EndOfBibitem
\bibitem[Khokhlov and Belov(2020)Khokhlov, and Belov]{Khokhlov2020b}
Khokhlov,~D.; Belov,~A. {Ab Initio Study of Low-Lying Excited States of
  Carotenoid-Derived Polyenes}. \emph{Journal of Physical Chemistry A}
  \textbf{2020}, \emph{124}, 5790--5803\relax
\mciteBstWouldAddEndPuncttrue
\mciteSetBstMidEndSepPunct{\mcitedefaultmidpunct}
{\mcitedefaultendpunct}{\mcitedefaultseppunct}\relax
\EndOfBibitem
\bibitem[Valentine \latin{et~al.}(2020)Valentine, Manawadu, and
  Barford]{Valentine20}
Valentine,~D.~J.; Manawadu,~D.; Barford,~W. {Higher energy triplet-pair states
  in polyenes and their role in intramolecular singlet fission}. \emph{Phys.
  Rev. B} \textbf{2020}, \emph{102}, 125107\relax
\mciteBstWouldAddEndPuncttrue
\mciteSetBstMidEndSepPunct{\mcitedefaultmidpunct}
{\mcitedefaultendpunct}{\mcitedefaultseppunct}\relax
\EndOfBibitem
\bibitem[Manawadu \latin{et~al.}(2022)Manawadu, Valentine, Marcus, and
  Barford]{Manawadu2022}
Manawadu,~D.; Valentine,~D.~J.; Marcus,~M.; Barford,~W. {Singlet Triplet-Pair
  Production and Possible Singlet-Fission in Carotenoids}. \emph{The Journal of
  Physical Chemistry Letters} \textbf{2022}, \emph{13}, 1344--1349\relax
\mciteBstWouldAddEndPuncttrue
\mciteSetBstMidEndSepPunct{\mcitedefaultmidpunct}
{\mcitedefaultendpunct}{\mcitedefaultseppunct}\relax
\EndOfBibitem
\bibitem[Barford(2022)]{Barford2022}
Barford,~W. {Theory of the dark state of polyenes and carotenoids}. \emph{Phys.
  Rev. B} \textbf{2022}, \emph{106}, 35201\relax
\mciteBstWouldAddEndPuncttrue
\mciteSetBstMidEndSepPunct{\mcitedefaultmidpunct}
{\mcitedefaultendpunct}{\mcitedefaultseppunct}\relax
\EndOfBibitem
\bibitem[Manawadu \latin{et~al.}(2022)Manawadu, Valentine, and
  Barford]{Manawadu2022a}
Manawadu,~D.; Valentine,~D.~J.; Barford,~W. {Dynamical simulations of
  carotenoid photoexcited states using density matrix renormalization group
  techniques}. 2022; \url{https://arxiv.org/abs/2211.02022}\relax
\mciteBstWouldAddEndPuncttrue
\mciteSetBstMidEndSepPunct{\mcitedefaultmidpunct}
{\mcitedefaultendpunct}{\mcitedefaultseppunct}\relax
\EndOfBibitem
\bibitem[Daley \latin{et~al.}(2004)Daley, Kollath, Schollw{\"{o}}ck, and
  Vidal]{Daley2004a}
Daley,~A.~J.; Kollath,~C.; Schollw{\"{o}}ck,~U.; Vidal,~G. {Time-dependent
  density-matrix renormalization-group using adaptive effective Hilbert
  spaces}. \emph{Journal of Statistical Mechanics: Theory and Experiment}
  \textbf{2004}, \emph{2004}, P04005\relax
\mciteBstWouldAddEndPuncttrue
\mciteSetBstMidEndSepPunct{\mcitedefaultmidpunct}
{\mcitedefaultendpunct}{\mcitedefaultseppunct}\relax
\EndOfBibitem
\bibitem[White and Feiguin(2004)White, and Feiguin]{White2004b}
White,~S.~R.; Feiguin,~A.~E. {Real-time evolution using the density matrix
  renormalization group}. \emph{Physical Review Letters} \textbf{2004},
  \emph{93}, 076401\relax
\mciteBstWouldAddEndPuncttrue
\mciteSetBstMidEndSepPunct{\mcitedefaultmidpunct}
{\mcitedefaultendpunct}{\mcitedefaultseppunct}\relax
\EndOfBibitem
\bibitem[Wasielewski \latin{et~al.}(1989)Wasielewski, Johnson, Bradford, and
  Kispert]{Wasielewski1989}
Wasielewski,~M.~R.; Johnson,~D.~G.; Bradford,~E.~G.; Kispert,~L.~D.
  {Temperature dependence of the lowest excited singlet-state lifetime of
  all-trans-$\beta$-carotene and fully deuterated all-trans-$\beta$-carotene}.
  \emph{The Journal of Chemical Physics} \textbf{1989}, \emph{91},
  6691--6697\relax
\mciteBstWouldAddEndPuncttrue
\mciteSetBstMidEndSepPunct{\mcitedefaultmidpunct}
{\mcitedefaultendpunct}{\mcitedefaultseppunct}\relax
\EndOfBibitem
\bibitem[Pol{\'{i}}vka \latin{et~al.}(1999)Pol{\'{i}}vka, Herek, Zigmantas,
  {\AA}kerlund, and Sundstr{\"{o}}m]{Polivka1999}
Pol{\'{i}}vka,~T.; Herek,~J.~L.; Zigmantas,~D.; {\AA}kerlund,~H.~E.;
  Sundstr{\"{o}}m,~V. {Direct observation of the (forbidden) $S_1$ state in
  carotenoids}. \emph{Proceedings of the National Academy of Sciences of the
  United States of America} \textbf{1999}, \emph{96}, 4914--4917\relax
\mciteBstWouldAddEndPuncttrue
\mciteSetBstMidEndSepPunct{\mcitedefaultmidpunct}
{\mcitedefaultendpunct}{\mcitedefaultseppunct}\relax
\EndOfBibitem
\bibitem[Gradinaru \latin{et~al.}(2001)Gradinaru, Kennis, Papagiannakis, {Van
  Stokkum}, Cogdell, Fleming, Niederman, and {Van Grondelle}]{Gradinaru2001b}
Gradinaru,~C.~C.; Kennis,~J.~T.; Papagiannakis,~E.; {Van Stokkum},~I.~H.;
  Cogdell,~R.~J.; Fleming,~G.~R.; Niederman,~R.~A.; {Van Grondelle},~R. {An
  unusual pathway of excitation energy deactivation in carotenoids:
  Singlet-to-triplet conversion on an ultrafast timescale in a photosynthetic
  antenna}. \emph{Proceedings of the National Academy of Sciences of the United
  States of America} \textbf{2001}, \emph{98}, 2364--2369\relax
\mciteBstWouldAddEndPuncttrue
\mciteSetBstMidEndSepPunct{\mcitedefaultmidpunct}
{\mcitedefaultendpunct}{\mcitedefaultseppunct}\relax
\EndOfBibitem
\bibitem[Fujii \latin{et~al.}(2003)Fujii, Inaba, Watanabe, Koyama, and
  Zhang]{Fujii2003}
Fujii,~R.; Inaba,~T.; Watanabe,~Y.; Koyama,~Y.; Zhang,~J.~P. {Two different
  pathways of internal conversion in carotenoids depending on the length of the
  conjugated chain}. \emph{Chemical Physics Letters} \textbf{2003}, \emph{369},
  165--172\relax
\mciteBstWouldAddEndPuncttrue
\mciteSetBstMidEndSepPunct{\mcitedefaultmidpunct}
{\mcitedefaultendpunct}{\mcitedefaultseppunct}\relax
\EndOfBibitem
\bibitem[Hallberg(1995)]{PhysRevB.52.R9827}
Hallberg,~K.~A. {Density-matrix algorithm for the calculation of dynamical
  properties of low-dimensional systems}. \emph{Phys. Rev. B} \textbf{1995},
  \emph{52}, R9827--R9830\relax
\mciteBstWouldAddEndPuncttrue
\mciteSetBstMidEndSepPunct{\mcitedefaultmidpunct}
{\mcitedefaultendpunct}{\mcitedefaultseppunct}\relax
\EndOfBibitem
\bibitem[K{\"{u}}hner and White(1999)K{\"{u}}hner, and White]{Kuhner1999a}
K{\"{u}}hner,~T.~D.; White,~S.~R. {Dynamical correlation functions using the
  density matrix renormalization group}. \emph{Physical Review B - Condensed
  Matter and Materials Physics} \textbf{1999}, \emph{60}, 335--343\relax
\mciteBstWouldAddEndPuncttrue
\mciteSetBstMidEndSepPunct{\mcitedefaultmidpunct}
{\mcitedefaultendpunct}{\mcitedefaultseppunct}\relax
\EndOfBibitem
\bibitem[Polak \latin{et~al.}()Polak, Musser, Sutherland, Auty, Branchi,
  Dzurnak, Chidgey, Cerullo, Hunter, and Clark]{Polak2019}
Polak,~D.~W.; Musser,~A.~J.; Sutherland,~G.~A.; Auty,~A.; Branchi,~F.;
  Dzurnak,~B.; Chidgey,~J.; Cerullo,~G.; Hunter,~C.~N.; Clark,~J. {Band-edge
  Excitation of Carotenoids Removes $S^*$ Revealing Triplet-pair Contributions
  to the $S_1$ Absorption Spectrum}. \emph{arXiv:1901.04900.} \relax
\mciteBstWouldAddEndPunctfalse
\mciteSetBstMidEndSepPunct{\mcitedefaultmidpunct}
{}{\mcitedefaultseppunct}\relax
\EndOfBibitem
\bibitem[Zhang \latin{et~al.}(2001)Zhang, Skibsted, Fujii, and
  Koyama]{Zhang2001a}
Zhang,~J.-P.; Skibsted,~L.~H.; Fujii,~R.; Koyama,~Y. {Transient Absorption from
  the $1B_u^+$ State of All-trans-$\beta$-carotene Newly Identified in the
  Near-infrared Region}. \emph{Photochemistry and Photobiology} \textbf{2001},
  \emph{73}, 219\relax
\mciteBstWouldAddEndPuncttrue
\mciteSetBstMidEndSepPunct{\mcitedefaultmidpunct}
{\mcitedefaultendpunct}{\mcitedefaultseppunct}\relax
\EndOfBibitem
\bibitem[Smith and Michl(2010)Smith, and Michl]{Smith2010}
Smith,~M.~B.; Michl,~J. {Singlet Fission}. \emph{Chemical Reviews}
  \textbf{2010}, \emph{110}, 6891--6936\relax
\mciteBstWouldAddEndPuncttrue
\mciteSetBstMidEndSepPunct{\mcitedefaultmidpunct}
{\mcitedefaultendpunct}{\mcitedefaultseppunct}\relax
\EndOfBibitem
\bibitem[Essler(2005)]{Essler2005}
Essler,~F. H.~L. \emph{{The one-dimensional Hubbard model}}; Cambridge
  University Press: Cambridge, 2005\relax
\mciteBstWouldAddEndPuncttrue
\mciteSetBstMidEndSepPunct{\mcitedefaultmidpunct}
{\mcitedefaultendpunct}{\mcitedefaultseppunct}\relax
\EndOfBibitem
\bibitem[Dixit and Mazumdar(1984)Dixit, and Mazumdar]{Dixit1984}
Dixit,~S.~N.; Mazumdar,~S. {Electron-Electron Interaction Effects on Peierls
  Dimerization in a Half-Filled Band}. \emph{Physical Review B} \textbf{1984},
  \emph{29}, 1824--1839\relax
\mciteBstWouldAddEndPuncttrue
\mciteSetBstMidEndSepPunct{\mcitedefaultmidpunct}
{\mcitedefaultendpunct}{\mcitedefaultseppunct}\relax
\EndOfBibitem
\bibitem[Musser \latin{et~al.}(2015)Musser, Maiuri, Brida, Cerullo, Friend, and
  Clark]{Musser2015}
Musser,~A.~J.; Maiuri,~M.; Brida,~D.; Cerullo,~G.; Friend,~R.~H.; Clark,~J.
  {The Nature of Singlet Exciton Fission in Carotenoid Aggregates}.
  \emph{Journal of the American Chemical Society} \textbf{2015}, \emph{137},
  5130--5139\relax
\mciteBstWouldAddEndPuncttrue
\mciteSetBstMidEndSepPunct{\mcitedefaultmidpunct}
{\mcitedefaultendpunct}{\mcitedefaultseppunct}\relax
\EndOfBibitem
\end{mcitethebibliography}


\providecommand{\latin}[1]{#1}
\makeatletter
\providecommand{\doi}
  {\begingroup\let\do\@makeother\dospecials
  \catcode`\{=1 \catcode`\}=2 \doi@aux}
\providecommand{\doi@aux}[1]{\endgroup\texttt{#1}}
\makeatother
\providecommand*\mcitethebibliography{\thebibliography}
\csname @ifundefined\endcsname{endmcitethebibliography}
  {\let\endmcitethebibliography\endthebibliography}{}
\begin{mcitethebibliography}{3}
\providecommand*\natexlab[1]{#1}
\providecommand*\mciteSetBstSublistMode[1]{}
\providecommand*\mciteSetBstMaxWidthForm[2]{}
\providecommand*\mciteBstWouldAddEndPuncttrue
  {\def\EndOfBibitem{\unskip.}}
\providecommand*\mciteBstWouldAddEndPunctfalse
  {\let\EndOfBibitem\relax}
\providecommand*\mciteSetBstMidEndSepPunct[3]{}
\providecommand*\mciteSetBstSublistLabelBeginEnd[3]{}
\providecommand*\EndOfBibitem{}
\mciteSetBstSublistMode{f}
\mciteSetBstMaxWidthForm{subitem}{(\alph{mcitesubitemcount})}
\mciteSetBstSublistLabelBeginEnd
  {\mcitemaxwidthsubitemform\space}
  {\relax}
  {\relax}

\bibitem[Taffet \latin{et~al.}(2019)Taffet, Lee, Toa, Pace, Rumbles, Southall,
  Cogdell, and Scholes]{Taffet2019e}
Taffet,~E.~J.; Lee,~B.~G.; Toa,~Z. S.~D.; Pace,~N.; Rumbles,~G.; Southall,~J.;
  Cogdell,~R.~J.; Scholes,~G.~D. {Carotenoid Nuclear Reorganization and
  Interplay of Bright and Dark Excited States}. \emph{The Journal of Physical
  Chemistry B} \textbf{2019}, \emph{123}, 8628--8643\relax
\mciteBstWouldAddEndPuncttrue
\mciteSetBstMidEndSepPunct{\mcitedefaultmidpunct}
{\mcitedefaultendpunct}{\mcitedefaultseppunct}\relax
\EndOfBibitem
\bibitem[Manawadu \latin{et~al.}(2022)Manawadu, Valentine, and
  Barford]{Manawadu2022a}
Manawadu,~D.; Valentine,~D.~J.; Barford,~W. {Dynamical simulations of
  carotenoid photoexcited states using density matrix renormalization group
  techniques}. 2022; \url{https://arxiv.org/abs/2211.02022}\relax
\mciteBstWouldAddEndPuncttrue
\mciteSetBstMidEndSepPunct{\mcitedefaultmidpunct}
{\mcitedefaultendpunct}{\mcitedefaultseppunct}\relax
\EndOfBibitem
\end{mcitethebibliography}

\end{document}


\pagebreak

\section{Parametrization of the UV-Peierls Hamiltonian}\label{Se:1}

\begin{figure}[h!]
 \includegraphics{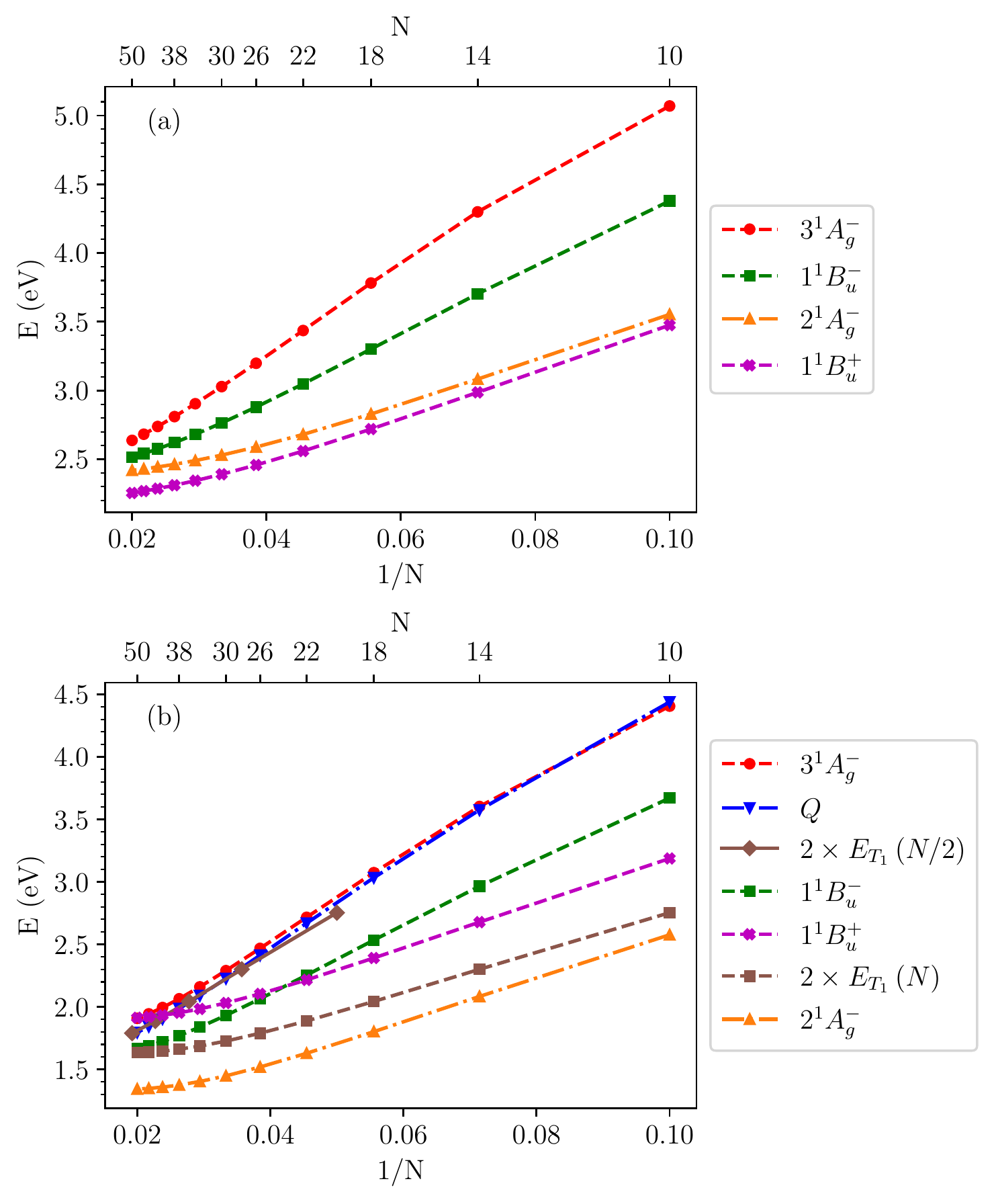}
  \caption{Vertical (a) and relaxed (b) singlet excitation energies of UV-Peierls model with $U = 7.25$ eV and $V = 3.25$ eV, found by solving eq (5) of the main paper. $N$ is the number of conjugated carbon atoms of the system. The vertical energy gaps of $\sim 0.1$ eV between $1^1 B_u^+$ (magenta) and $2^1 A_g^-$ (orange) for $18 \le N \le 26$ agree with the corresponding excitation energies reported in ref \cite{Taffet2019e}.}
  \label{Figure1}
\end{figure}

Figure \ref{Figure1} illustrates the diabatic vertical and relaxed excitation energies of the UV-Peierls model parametrized for direct internal conversion from the $1^1 B_u^+$ state to the $2^1 A_g^-$ state. The parametrization is performed such that the vertical excitation energy gap between the $1^1 B_u^+$ to the $2^1 A_g^-$ states reproduces the excitation energies reported in Table 2 of ref \citen{Taffet2019e}.

\section{Parametrization of the symmetry breaking Hamiltonian, $\hat{H}_\epsilon$}\label{Se:1}

As described in ref \citen{Manawadu2022a}, $\hat{H}_\epsilon$ is optimized under the constraint $|\epsilon_n| < \epsilon_{\textrm{max}}$ such that the ground state $\pi$-electron density on C-atom $n$ reproduces the Mulliken charge densities of the $\pi$-system found \emph{via} ab initio density functional theory (DFT) calculations.
The optimized $\hat{H}_\epsilon$ is given in Table \ref{Heps}. The cut-off $\epsilon_{\textrm{max}}=1.0$ is chosen such that $\Psi(t=0)$ retains sufficient $1^1 B_u^+$ character while accurately reproducing the DFT densities with a coefficient of variation $r^2(\boldsymbol{\epsilon}) = 0.92$.

\begin{table}[h!]
    \centering
    \begin{tabular}{|r|r|r|r|r|r|}
    \hline
    	\multicolumn{2}{|c|}{~}  	& \multicolumn{2}{|c|}{$V=2.75$ eV} & \multicolumn{2}{|c|}{$V=3.25$ eV}  \\ \hline
        Carbon site, $n$ & Mulliken charges (q) & $\epsilon_n$ (eV) & $\expval{\hat{N}_n -1}$ & $\epsilon_n$ (eV) & $\expval{\hat{N}_n -1}$ \\ \hline
        1 & 0.14 & -1.00 & 0.17&-1.00 & 0.17 \\ \hline
        2 & -0.18 & 0.81 & -0.14 &0.56 & -0.14 \\ \hline
        3 & -0.05 & 1.00 & 0.05 &1.00 & 0.06 \\ \hline
        4 & -0.18 & 1.00 & -0.10 & 0.82 & -0.10 \\ \hline
        5 & 0.15 & -1.00 & 0.15 & -1.00 & 0.15 \\ \hline
        6 & -0.14 & 0.27 & -0.10 & 0.02 & -0.10 \\ \hline
        7 & -0.07 & 1.00 & 0.03 &1.00 & 0.03 \\ \hline
        8 & -0.16 & 1.00 & -0.09 &0.84 & -0.09 \\ \hline
        9 & 0.13 & -1.00 & 0.12 &-1.00 & 0.12 \\ \hline
        10 & -0.09 & 0.14 & -0.07 &-0.01 & -0.07 \\ \hline
        11 & -0.11 & 1.00 & -0.03 &1.00 & -0.04 \\ \hline
        12 & -0.10 & 1.00 & -0.02 & 1.00 & -0.03 \\ \hline
        13 & -0.11 & 0.32 & -0.09 & 0.09 & -0.09 \\ \hline
        14 & 0.14 & -1.00 & 0.13 & -1.00 & 0.13 \\ \hline
        15 & -0.18 & 1.00 & -0.09 &0.92 & -0.10 \\ \hline
        16 & -0.05 & 1.00 & 0.05 & 1.00 & 0.06 \\ \hline
        17 & -0.19 & 0.90 & -0.14 & 0.61 & -0.14 \\ \hline
        18 & 0.14 & -1.00 & 0.18 & -1.00 & 0.17 \\ \hline
    \end{tabular}
    \caption{The $\pi$-electron Mulliken charges from the \emph{ab-initio} DFT calculation, parameters for $\hat{H}_{\epsilon}$ found for $\epsilon_{\textrm{max}} =1.0$ eV, and the expectation values of number densities calculated from the parametrized $\hat{H}_{\epsilon}$.
    In order to maintain $\pi$-electron charge neutrality, each \emph{ab-initio} charge was increased by $0.05q$. The chemical formula of neurosporene is shown in Figure 1 of the main paper.}
    \label{Heps}
\end{table}
\newpage
\section{Probabilities that the adiabatic states, $S_1$, $S_2$ and $S_3$ occupy the diabatic states $2^1 A_g^-$, $1^1 B_u^+$ and $1^1 B_u^-$}\label{Se:2}

\begin{figure}[h!]
\centering
         \includegraphics{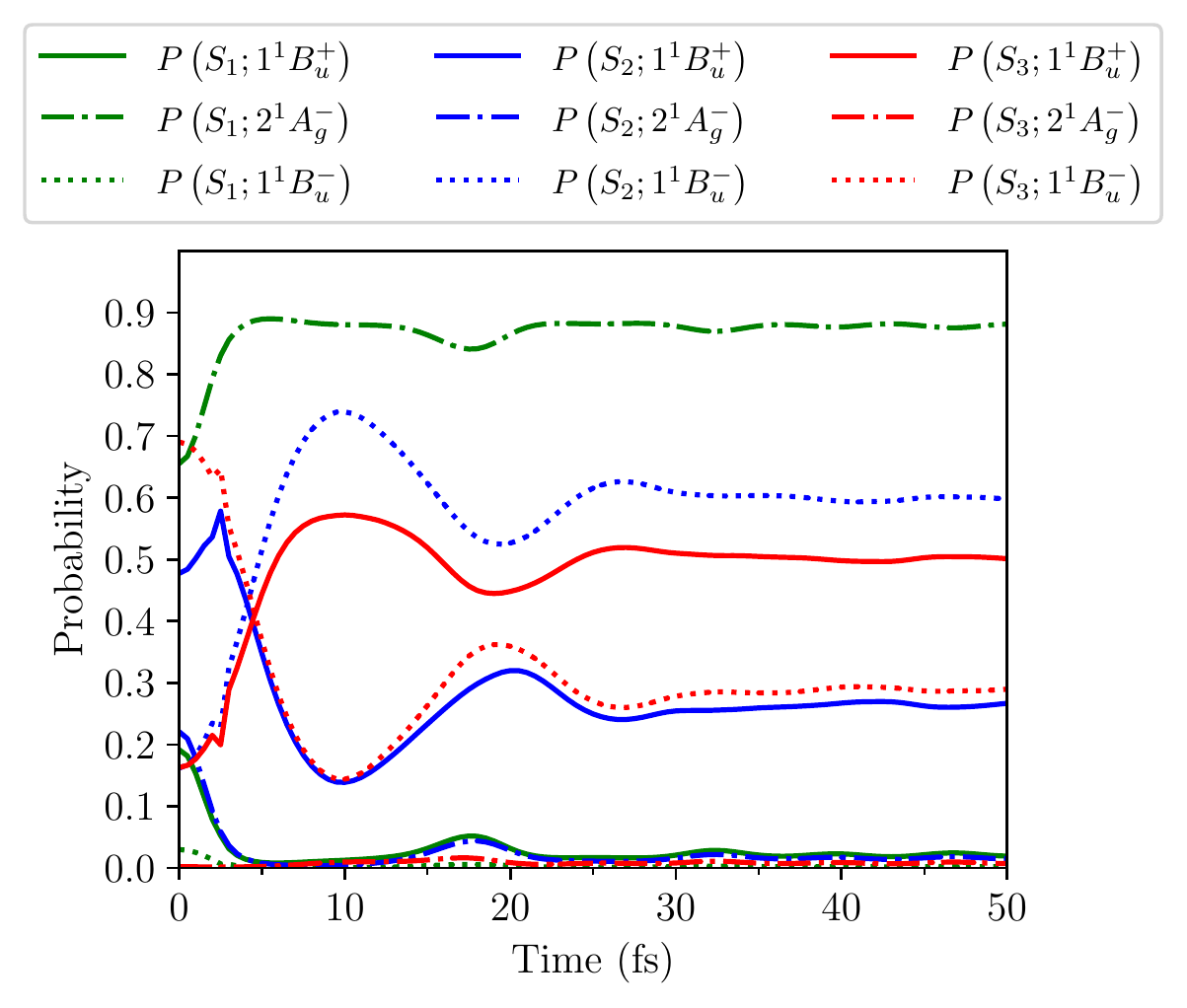}
  \caption{Probabilities as a function of time that the adiabatic states, $S_1$, $S_2$ and $S_3$, occupies the diabatic states, $2^1 A_g^-$, $1^1 B_u^+$ and $1^1 B_u^-$. Results are for neurosporene ($N=18$) with $V=2.75$ eV.}
   \label{Figure5}
\end{figure}

Figure \ref{Figure5} illustrates the probabilities that the adiabatic states $S_1$, $S_2$, and $S_3$ occupy the diabatic states $2^1 A_g^-$, $1^1 B_u^+$, and $1^1 B_u^-$. Adiabatic states are $\sim 90\%$ occupied by the diabatic states at all times. Using the probabilities that the triplet-pair states, $2^1 A_g^-$ and $1^1 B_u^-$, occupy the adiabatic states, $S_1$ and $S_2$, the `classical' total triplet-pair yield can be calculated via eq (7) of the main paper.



\bibliography{D.Phil-Writing-Experimental_JPCA}

\vfill